\documentclass{article}
\usepackage[utf8]{inputenc}

\usepackage{graphics}
\usepackage{adjustbox}
\usepackage{caption, subcaption}

\usepackage{hyperref}
\usepackage[dvipsnames]{xcolor}

\title{Online voluntary mentoring: Optimising the assignment of students and mentors}
\author{P\'eter Bir\'o and M\'arton Gyetvai\\ Institute of Economics, CERS, and\\ Corvinus University of Budapest\\ \texttt{\{peter.biro,marton.gyetvai\}@krtk.mta.hu}}
\begin{document}

\maketitle

\begin{abstract}
\noindent After the closure of the schools in Hungary from March 2020 due to the pandemic, many students were left at home with no or not enough parental help for studying, and in the meantime some people had more free time and willingness to help others in need during the lockdown. In this paper we describe the optimisation aspects of a joint NGO project for allocating voluntary mentors to students using a web-based coordination mechanism. The goal of the project has been to form optimal pairs and study groups by taking into the preferences and the constraints of the participants. In this paper we present the optimisation concept, and the integer programming techniques used for solving the allocation problems. Furthermore, we conducted computation simulations on real and generated data for evaluate the performance of this dynamic matching scheme under different parameter settings. 
\vspace*{0.1cm}

\end{abstract}

\section{Introduction}

In Hungary, after the escalation of the COVID-19 virus, the government announced the closure of all schools on 13 March 2020 (Friday evening) with effect from 16 March, and they requested all the schools (both primary and secondary schools for students at ages between 6 and 18) to start online/distance education immediately. There was no central recommendation about the technology and methodology used, so this was decided mainly by the board of each school using a wide range of online platforms (e.g. Google Classroom, MS Teams, etc) or just sending the weekly assignments by post. 

A large number of students had difficulties to follow the distance education, partly because of the lack of equipment or internet connection, but also because the parents were not be able to help them at home, due to their lack of knowledge in special subjects or just because of being at work (e.g., including those parents working intensively in health care). At the same time, many people, especially the elderly ones had to stay at home with spare time. Furthermore, in Hungary there is also a scheme for secondary school students for voluntary work, which they can get credit for (that counts extra points at the centralised university admission), and very few possibilities remained for such services under the strict social distancing rules imposed. Therefore, there was both a large need for mentoring, and also a significant amount of potential mentors, both elderly and young.

A project was proposed in early April and then officially started with the opening of a web-application (\texttt{onkentesmentoralas.hu}) in early May in a cooperation of three parties. On behalf of the Institute of Economics of CERS, the Mechanism Design and the Education Economics research groups offered their help in designing a mechanism for allocating students to mentors. The Hungarian Reformat Church Aid is a humanitarian organisation, which had a link to a governmental action group devoted to coordinate the voluntary help in Hungary. The third party in the project was \#school, a private company providing an online teaching platform, that became used by 100k registered users soon after the online education started. The latter two parties have had a related collaboration in the past, where they organised the mentoring of highly disadvantaged children.

The design approach was rather complex, taking into account the preferences of both sides, and also allowing the formation of study groups, besides the mentor-student pairs. The basic requirement of creating a pair is to have a subject (e.g. Math at year 7) that is both requested by a student and offered by a mentor. However, the students can request several subjects listing them according to their preferences and whether they are willing to study in groups or only in pairs. The mentors can specify their preferences over the subjects they offer to teach, whether they are willing to supervise groups (or only individuals), and they can also set preferences over some characteristics of the students, such as their age, their performance level in that subject measured by their grades, and their social status (i.e. whether the mentor prefer to teach highly disadvantaged students). 

In the optimisation we carefully considered several potential goals and we implemented a combination of them, such as maximising the number of students matched, the overall volume of the teaching hours, the preferences of both the students and mentors, and the coherence of the study groups. After implementing our approach we conducted matching runs on a week from early May until the end of the academic year, that is mid-June in Hungary.

The first period of this application was very short, and we had an unexpectedly high number of mentors volunteered, but also an unexpectedly low number of students registered. Therefore the allocation problem was rather straightforward. Nevertheless, we used the data from this early period to generate instances that are realistic, and we present the simulations conducted on that data in this paper.

The pandemic situation has became critical again in Hungary by October due to the second wave of infections, so on 6 November 2020 all the secondary schools and universities have been closed again, and thus we open again our allocation service to link volunteering mentors with students in need. %\PB{Itt meg erdemes lesz megkerdezni beadas elott, hogy vannak-e uj adatok, amiket esetleg riportalni lehetne}

The pandemic situation has been similar around the world, and especially critical since April in the USA, an early report on the American schools' responses to COVID-19 can be found in \cite{Harris2020}.

In this paper we describe the optimisation aspects of our aforementioned joint NGO project of allocating voluntary mentors to students with presenting also computational simulations on realistic instances. We believe that this paper provides an interesting case study with an advanced OR solution that could be used everywhere in the world to help allocating the volunteers to people in need in an efficient and fair way.

\subsection{Related literature}

Matching problems under preferences in two-sided matching markets have been widely studied both in mathematics, computer science and economics, see e.g.\ a recent book on the algorithmic aspects of this topic by Manlove~\cite{Manlove2013book}. Beside the theoretical studies, practical applications have been designed and implemented in many areas, see a recent survey on this \cite{biro2017applications}. 

When both sides of the application have preferences, then the concept of stable matchings was proposed in the seminal paper of Gale and Shapley \cite{gale1962college} and has been used since in many applications, such as resident allocation, college admission and school choice. 
However, there can be some special features that can make the stable matching problem computationally hard to solve. In this case one robust approach to tackle these problems is (mixed) integer linear programming, that has been used recently for the hospital--resident problem with couples \cite{BMMcB2014}, ties \cite{KM2014,Delormeetal2019}, college admissions with lower and common quotas \cite{ABMcB2016}, and stable project allocation under distributional constraints \cite{ABSz2018}. In this paper we also use MILP technique for solving the underlying optimisation problem.

There are also many application, where preference of one or both sides do matter, but the solution is not necessarily stable or fair in some sense, but rather optimal. Examples are the allocation of papers to reviewers \cite{garg2010assigning}, course allocation \cite{budish2017course}, or arranged marriages  \cite{Cao2010}. Scheduling problems are also closely related, see a paper linking the two lines of researches \cite{Biro2014}.

In our application of allocating mentors to students we can observe a dynamic nature, so the literature on online matching is also related. Natural applications for online matching with preferences are deceased organ allocation \cite{Mattei2018}, \cite{Agarwal2020}, allocation in social housing \cite{leshno2019dynamic}, \cite{bloch2020matching}, electric vehicle charging \cite{Gerding2019}, or lending decisions \cite{So2016}.

There are also applications, which are dynamic in nature, but instead of online matching protocols, batch allocations are also used. An important example is refugee allocation, where preferences of one or both sides may be taken into account, together with some objective goals of maximising the likelihood of successful settlement of the refugee families, see \cite{trapp2018placement}, \cite{andersson2018dynamic}, \cite{bansak2018improving}. Similar approaches are used in the allocation of foodbanks \cite{prendergast2016allocation}.

Finally, an important example for an application where optimisation is used for the allocation are the kidney exchange programmes (KEPs), where kidney patients with incompatible donors may exchange their willing donor among themselves. Seminal work on IP models for KEP's is presented in \cite{Abraham2007} and \cite{Roth2007}, a recent survey is \cite{AshlagiRoth2020}, and the European optimisation practices are summarised in \cite{biroetal2019b}. It is interesting to note that online matching is used in the US, partly because of the competition in between multiple national programmes \cite{agarwal2019market}. However, in Europe the national programmes use batch allocations, by conducting the matching runs in 3-4 months regular intervals \cite{biroetal2019a}.

\subsection{Our contributions}

We describe the allocation mechanism that we designed with our partners and implemented in the applications. The design is complex, it takes into account the preferences of both sides, and also some objective factors. The main novelty and challenge in our solution concept is that besides mentor-student pairs we also seek to form study groups, that makes the underlying optimisation problem more elaborate. 

Our main theoretical contribution is an IP formulation that accommodates all the complex constraints and objectives of our model. 

These results are complemented by computational experiments, where the generation of the instances is based on the real data that we collected in the first period of the application. In our simulation we analyse the effect of some optimisation policy decisions with regard to various performance measures. We also test the effects of having shorter or longer matching periods, and the possibility of giving priority based on the waiting time. The results of the simulation have been used to refine the design and optimisation policy of the application.

\section{Preliminaries}

When describing the formal model, first we give a general description, then we specify the variables for the solution, and finally we describe the input data for the preferences, priorities and further objective factors.

\subsection*{General description}

We have a set of students $A=\{a_1, a_2, \dots , a_k\}$, mentors $B=\{b_1, b_2, \dots , b_l\}$, and subjects $S=\{s_1,s_2, \dots, s_m\}$, the latter being specified with the year as well (e.g., year 10 - Chemistry). We would like to form pairs (each consisting of one student and one mentor) and groups (each consisting of a set of students and one mentor) so that the online mentoring is conducted in these pairs and groups in weekly periods. For every pair and group formed we also specify the subjects that they are going to study and the amount of time that they are supposed to spend with each subject during a week.

As the input of the problem the students give the list of subjects that they need mentoring in a preference order, specifying also the amount of time they wish to spend with a mentor per week (integer number between 1 and 3) for each subject. So a student may ask 2 hours of mentoring in Maths and 1 hour in Physics with higher preference for the former. The students can give information on why they need mentoring (e.g., being a child of a single mother who works in a hospital as a nurse), and about their objective circumstances with regard to their social background, such as the number of children/parent at home, etc, that can result in additional priority for them. Furthermore, they also provide information on the class they attend and whether they have ongoing online education from that subject, if they are weak, medium or good students (by giving their final grade in the last semester), and whether they are willing to accept mentoring in groups or only in pairs. Finally, we ask what equipment they have at home (PC, tablet, smartphone). 

Regarding the mentors, we ask in which subjects they are willing to mentor students (including the years), and whether they have preference over the subjects, and some characteristic of the students, such as their age, how strong the students are in the subject, and their social status (some mentors can be especially keen to teach socially disadvantaged or weak-performing students, some might not). We ask the total number of hours that they wish to spend with mentoring per week and whether they are willing to do mentoring in groups.

The organisers of this project decided that some of the above mentioned priorities will only apply for such mentors that express their agreement. For instance, every mentor can tell whether she wishes to teach highly disadvantaged students, and if they say no then no priority is given for those students when considering to allocate them to this mentor. However, if the mentor states that she would be happy to teach highly disadvantaged students then a high weight is added. If the mentor is ignorant about this aspect, then a small weight is added for this priority factor. The same applies for the criterion whether the student is a weak student with low grades from the last semester. If a mentor expresses that she does not want to teach weak students then no priority is given for these pairs, but if she wishes to teach weak students then extra weight is given. Thus we take the preferences of the mentors into accounts for some of these controversial priorities, but for some others, e.g., number of children/parents in the family, we always give an extra priority. In this way the pairs and group resulted are more likely to be mutually satisfying.

\subsection*{Notations: basic elements of the solution}

As the basic building block of the optimisation model we have the set of possible paired mentoring activities $E$, where each \emph{activity} $e$ consists of a triple $e=(a_i,b_j,s_k)$. Activity $e$ is possible if student $a_i$ requested subject $s_k$ and mentor $b_j$ also offered $s_k$.

Let $S(a_i)$ and $S(b_j)$ denote the set of requested subjects by student $a_i$ and offered subjects by mentor $b_j$, respectively. Furthermore, student $a_i$ requested $q(a_i,s_k)$ hours per week in subject $s_k$ and mentor $b_j$ offered $Q(b_j)$ hours per week in total. We write that $(a_i,s_k)\in e$ if there is $b_j\in B$ such that $e=(a_i,b_j,s_k)$, and similarly, we write that $(b_j,s_k)\in e$ if there is $a_i\in A$ such that $e=(a_i,b_j,s_k)$.

In the description of a solution, let $x_e$ denote the amount of hours scheduled for activity $e$ in a pair, and let $y_e$ be a binary variable denoting whether activity $e$ is performed in a pair, i.e. $y_e=1 \iff x_e>0$. In our practical application we restrict $x_e\in\{0,1,2,3\}$.\footnote{One can consider to set these variables to be continuous in the interval $[0,3]$. However, because of the study groups, we might get fractional values in an optimal solution, which is not allowed in practice.}  Let $P$ denote the set of pairs formed, i.e., $P=\{e\in E: y_e=1\}$.

Besides pairs, we also allow the formation of groups, whose final set in the solution will be denoted by $G$. Each \emph{group} $g\in G$ consists of a triple $g=(A^g, b_j, s_k)$, where $A^g\subset A$ is the subset of students involved in group $g$ with mentor $b_j$ and subject $s_k$. Here $x_g$ denotes the amount of time scheduled for group $g$, $x_g\in\{2,3\}$, if $g$ is formed. (Note that $x_g$ is not a variable in our MILP model, we use that only to describe our solution.)

In our model and application, for simplicity, we assume that every mentor can have at most five groups to supervise in a subject. Thus, for every mentor $b_j$ who is willing to teach groups in subject $s_k$, we create five \emph{potential groups} $g_j^{k,1}$, $g_j^{k,2}$,  $\dots$, $g_j^{k,5}$ with capacity $c_j^{k}$ each. We introduce a binary variable $y_j^{k,t}$ to denote whether the potential group $g_j^{k,t}$ is realised, in which case there are at least two and at most $c_j^k$ students involved. Let $x_j^{k,t}$ denote the number of hours allocated for its weekly operation, where $x_j^{k,t}\in\{0,2,3\}$. Note that $y_j^{k,t}$ and $x_j^{k,t}$ are the variables of our mixed integer programming formulation.

For every activity $e=(a_i,b_j,s_k)$, we create a binary indicator variable $y_e^t$ denoting whether this mentoring activity is performed in potential group $q_j^{k,t}$. Therefore the set of students involved in this group will be $A^g=\{a_i\in A: e=(a_i,b_j,s_k), y_e^t=1\}$. The following formula summarises the feasibility condition for realising potential group $q_j^{k,t}$.

$$y_j^{k,t}=1 \iff 2\leq \sum_{e: (b_j,s_k)\in e} y_e^t \leq c_j^k$$

For example, a potential group $g_j^{k,1}$ for mentor $b_j$ can be on 7-year Maths ($s_k$) for at most 5 students ($c_j^k=5$). If this group is realised ($y_j^{k,1}=1$), a mentoring group $g=(A^g,b_j,s_k)$ will be formed where $2\leq |A^g|\leq 5$ with $x_g=x_j^{k,1}$ hours per week. 

Regarding the personal mentoring hours in a group, as students in a group may have different requests, we estimate the actual mentoring hours by variable $x_e^t$, where we assume that $x_e^t\leq q_j^{k,t}$ and $x_e^t\leq q(a_i,s_k)$, where $e=(a_i,b_j,s_k)$. 

\subsection*{Preferences, priorities and further objective factors}

Here we describe the input more formally listing the information provided by the users with regard to their attributes and preferences.

\subsubsection*{Students' attributes and preferences}

For every student $a_i$, we collect the following attributes and preferences:\\
\begin{itemize}
    \item $year_i\in \{1, 2, \dots 12\}$: which year of study she attends.
    \item $class_i$: exact class she attends in the school (text).
    \item $PA_i$: preference list consisting of the subjects she requested mentoring (e.g., Math, Physics, History). Let $rank_i(s_k)$ denote the rank of $s_k$ in $PA_i$ for $s_k\in S(a_i)$.
    \item $RA_i$: number of requested hours for each requested subjects in the order of preferences. Note that in our LP model $q(a_i,s_k)$ denotes this constant for each subject $s_k\in S(a_i)$.
    \item $grades_i$: her grades from the last semester in the subjects requested, let $gr(a_i,s_k)$ denote the grade of $a_i$ in subject $s_k$, that is a value in between 1 and 5 in Hungary (5 being the best, and 1 meaning Failed).
    \item $group_i\in \{0,1\}$: the value is 1 if $a_i$ is willing to accept mentoring in groups.
    \item $equipment_i\in \{0,1\}$: the value is 0 if she has a smartphone or tablet and the value is 1 is she (also) has a laptop.
    \item $help_i\in \{0,1,2\}$: self-reported neediness, whether $a_i$ has help at home (0= yes, 1= limited, 2= no)
\end{itemize}

When forming the groups, the similarities between the students can be important, therefore for those students willing to get mentoring in groups we define the following values for each pair of students $a_i$ and $a_{i'}$.

\begin{itemize}
    \item $sc_{i,i'}$: the value is 1 if they attend the very same class and 0 otherwise.
    \item $dr_{i,i'}^k=|q(a_i,s_k)-q(a_{i'},s_k)|$: the difference between the requested amounts of time in subject $s_k$.
    \item $dg_{i,i'}^k=|gr(a_i,s_k)-gr(a_{i'},s_k)|$: the difference between their last year's grades in subject $s_k$.
    \item $de_{i,i'}=|equipment_i-equipment_{i'}|$: this value is 0 if they have the same equipment and 1 if they have different ones.
\end{itemize}

Furthermore, for those students wishing to get priority, we also ask question about her social background and circumstances, based on which we award the following scores for student $a_i$.

\begin{itemize}
    \item $SD_i\in \{0,1,2,3\}$: a measure showing how socially disadvantaged is the student (3 being the maximum value).
    \item $NH_i$: an index for not enough help at home with a value between 0.5 and 2.5 showing how many children are for one parent at home. Thus its value is a half-integer between 0.5 and 2.5.
    \item $WS_i\in \{0,1,2,3\}$: showing how weak is the student based on her grades and repeated years.
    \item $CY_i\in\{0,1,2\}$: how critical the year is for the student, i.e, for last year of studies $=2$, and for one before the last year $=1$.
\end{itemize}

\subsubsection*{Mentors' preferences}

Here we describe what preferences the mentors provide on the subjects and students.

\begin{itemize}
\item $PM_j$: preference list of $b_j$ on the subjects she offers for mentoring (e.g., Math, Physics, History). Let $rank_j(s_k)$ denote the rank of $s_k$ in $PM_j$ for $s_k\in S(b_j)$. Note that this may be set differently for the three different age-categories, years 1-4, 5-8, 9-12.
\item $YM_j\in \{0,1,2\}$: most preferred age of the student for $b_j$ (0= Year 1-4, 1= Year 5-8, 2= Year 9-12). 
\item $DM_j\in \{0,1,3\}$: whether $b_j$ is willing to mentor socially disadvantaged students (0= rather not, 1= does not matter, 3= would be very keen).
\item $GPM_j$: grade-preference, $=N$ if no preference is given, $=W$ for weak, $=M$ for medium, and $=S$ for strong students. From this information, we create the following constants: $PM_j=0$ if $GPM_j=N$ and $PM_j=1$ if  $GPM_j\neq N$;  $SM_j=0$ if $GPM_J\in \{M,S\}$, $SM_j=1$ if $GPM_J=N$ and $SM_j=3$ if $GPM_J=W$; finally     $WM_j=0$ if $GPM_J=N$, $WM_j=1.5$ if $GPM_J=W$, $WM_j=3$ if $GPM_J=M$, and $WM_j=4.5$ if $GPM_J=S$. Here, $PM_j$ is an indicator whether $b_j$ has grade-preferences; $SM_j$ show how willing $b_j$ is to mentor a weak student; finally, $WM_j$ is the best average grade of the student according to $b_j$'s preference. 
\end{itemize}

\section{Optimisation with MILP technique}\label{sec:LP}

In this section we show how we can formulate our problem as a mixed integer linear program. 

First we describe the basic constraints for the feasibility of a solution. We summarise the feasibility requirements as follows. 

\smallskip
\textbf{Feasibility constraints:}

\begin{enumerate}
\item Only use mutually acceptable paired mentoring activities in the solution (i.e., the subject should be requested by the student and offered by the mentor linked).
\item Only those mentors can have groups and only those students can be assigned to groups who expressed their willingness to teach or study in groups, respectively.
\item Have at least two and at most a limited number ($c_j^k$) of students in each group.
\item We obey the weekly capacities of the mentors.
\item The mentoring hours of a pair in a subject should not exceed the requested amount by the student, and never be more than 3. The mentoring hours per week for an active group is either 2 or 3.
\item Every student can be mentored by at most one mentor in each subject that she requested.
\end{enumerate}

The first two conditions are automatically satisfied, since we only work with mentoring activities, where the subjects are mutually acceptable by both parties (i.e., requested by students and offered by mentors). Similarly, we only create potential groups for those mentors who are willing to teach in groups and we only have variables $y_e^{t}$ and $x_e^{t}$ for those students who are willing to study in group.  

Regarding the pairs, the connection between $x_e$ and $y_e$ can be established with the following formula.

\begin{equation}\label{xy_e}
    y_e\leq x_e \leq 3\cdot y_e \mbox{ for every } e\in E
\end{equation}

A similar formula is added for each potential group, but with the minimum number of hours being 2.

\begin{equation}\label{xy_g}
    2\cdot y_j^{k,t}\leq x_j^{k,t} \leq 3\cdot y_j^{k,t} \mbox{ for every } b_j\in B, s_k\in S(b_j), t\in \{1,2,3\}
\end{equation}

For the realisation of potential groups, we set the following conditions. 

\begin{equation}\label{xy_g}
2\cdot y_j^{k,t} \leq \sum_{e: (b_j,s_k)\in e} y_e^t \leq c_j^k\cdot y_j^{k,t}
\end{equation}

Every activity can be used either in a pair or in a group:

\begin{equation}\label{activity}
y_e + \sum_t y_e^t \leq 1 \mbox{ for every } e\in E
\end{equation}

The above constraint is also enforced by the more general requirement that every student can have at most one mentor in each subject:

\begin{equation}\label{subject}
\sum_{e: (a_i,s_k)\in e} \left ( y_e + \sum_t y_e^t \right ) \leq 1 \mbox{ for every } s_k\in S(a_i), a_i\in A
\end{equation}

For later usage in the objective function, we introduce some new variables $\beta_i^k$ indicating whether student $a_i$ is involved in mentoring activity in subject $s_k$ and $\gamma_i$ indicating whether student $a_i$ is involved in any mentoring activity. We link these variables with the basic variables as follows: 
$$\beta_i^k = \sum_{e: (a_i,s_k)\in e} \left (y_e + \sum_t y_e^t \right ) \mbox{ for every } s_k\in S(a_i), a_i\in A,$$
and let
$$\gamma_i\leq \sum_{s_k\in S(a_i)} \beta_i^k\leq M\cdot \gamma_i \mbox{ for every } a_i\in A,$$ where $M$ is a large enough number, e.g., the number of subjects any student can possibly request mentoring (5 in our case).

The weekly capacity of the mentors should not be exceeded:

\begin{equation}\label{mentor_kap}
\sum_{s_k\in S(b_j)} \left( \sum_{e: (b_j,s_k)\in e} x_e + \sum_{t=\{1,\dots, 5\}} x_j^{k,t}\right) \leq Q(b_j) \mbox{ for every } b_j\in B
\end{equation}

The mentoring hours of a pair in a subject should not exceed the requested amount by the student:

\begin{equation}\label{student_kap} 
x_e \leq q(a_i,s_k) \mbox{ for every } e\in E, (a_i,s_k)\in e
\end{equation}

When computing the volume of a solution, for groups we assume that the actual mentoring hours for a student ($x_e^t$) is upper bounded by her original request in this subject, and by the number of mentoring hours of the group. This can be formalised with the following constraints

\begin{equation}\label{student_act1} x_e^t \leq q(a_i,s_k) \mbox{ for every } e\in E, (a_i,s_k)\in e, t=\{1,\dots, 5\}
\end{equation}

\begin{equation}\label{student_act2} x_e^t \leq x_j^{k,t} \mbox{ for every } e\in E, (a_i,s_k)\in e, t=\{1,\dots, 5\}
\end{equation}

Now, we turn our attention to the objectives.

\subsection*{Possible objectives}
\begin{enumerate}
\item Maximising the number of students getting a mentor.
\item Maximising the number of pairs and groups realised with different weights for pairs and groups.
\item Maximising the number of mentoring hours realised with different weights for pairs and groups.
\item Satisfying the preferences of the students and mentors as well with regard to the subjects.  
\item When forming groups, we improve cohesion if we have students a) from the same class, b) with the same type of equipment, c) their former grades being as close as possible, and d) the number of scheduled hours for the group to be close to what the assigned students requested in that subject.
\item Giving priority to certain students in needs (depending on the preferences of the mentors for some criteria, such as social status and student past performance).
\item When the same mentor-student pair can be involved in paired mentoring activities in multiple subjects then this is preferable.
\end{enumerate}

We note that additionally one might wish to consider the genders of the mentor-student pairs, or whether they live in the same city, geographic area, or maybe if they attend the same school in the case of student mentors. 

Let $w^g$ denote the discount for a mentoring activity being realised in a group, as opposed to a pair. In our default setting, we use $w^g=0.7$, which means that each mentoring hour in a group counts 0.7 hour in a pair. This is a crucial parameter, since it highly affects the share of groups in the final solution. 

Now we explain the rational behind each of the above listed objectives and we formulate the corresponding linear terms for the objective function of the linear program. The relative weights of these terms in the objective function were up to the expert choices by the organisers of this application. Note that the relative weights sum up to 100, so they can be interpreted as percentages showing the importance of the objective criteria.

\begin{enumerate}
\item \textbf{Number of students allocated.} Allocating mentors to as many students as possible.\\
\emph{Rational pros:} We would like to involve as many students as possible in the mentoring, even if for one subject requested.\\
\emph{Rational cons:} If this objective is dominating then we would have many mentoring activities with 1 hour only that can be inefficient for both students and mentors.\\
\emph{Relative importance in the application:} 0 (i.e., this criterion was decided not to be considered as objective, only monitored)\\
\emph{Linear term:} With the usage of variables $\gamma_i$, we can simply express this objective as follows.
$$\sum_{a_i\in A} \gamma_i$$

\item \textbf{Number of pairs and groups created.} Maximising the number of pairs and groups realised with different weights for pairs and groups.\\
\emph{Rational pros:} We would like to create as many pairs and groups as possible, to create the links between the parties.\\
\emph{Rational cons:} If this objective is dominating then we would have many mentoring activities with 1 hour only for pairs and 2 hours for groups that can be inefficient for both students and mentors.\\
\emph{Relative importance in the application:} 0 (i.e., this criterion was decided not to be considered as objective, only monitored)\\
\emph{Linear term:} Similar to the formula for $\beta_i$, we express this objective as follows.
$$\sum_{e\in E} \left( y_e + \sum_{t} w^g\cdot y_e^t \right)$$

\item \textbf{Volume.} Maximising the number of mentoring hours realised with different weights for pairs and groups.\\
\emph{Rational pros:} The number of mentoring hours is the most important measure.\\
\emph{Rational cons:} The solution might be unbalanced, some students may get many hours while some others may get none.\\
\emph{Relative importance in the application:} 50\\
\emph{Linear term:} Similar to the previous formula we now use the hours rather than the indicator variables, resulting in the volume of the solution.

$$\sum_{e\in E} \left (x_e + \sum_{t} w^g\cdot x_e^t \right )$$

As the volume will be the main objective in our optimisation, we will set the relative weights for each mentoring activity accordingly. Let $w_e$ denote the final weight of activity $e$ in a pair and let denote $w_e^g$ is the final weight of an activity in a group. We suppose that $w_e^g=w_e\cdot w^g=w_e\cdot 0.7$ in our case. The final weight $w_e$ will be a sum of weights with respect to different objectives. The first objective is the volume, that we weighted 50 for all activities, so let $w_e^w=50$ for every activity $e$.

\item \textbf{Preferences.} Satisfying the preferences of the students with regard to the subjects, and the preferences of the mentors on the subjects and on the ages of students.\\
\emph{Rational:} The higher the need of the student for a subject the more important for her to get help, and the preference of the mentor should also be taken into account for the subject and the age of the student supervised.\\
\emph{Relative importance in the application:} 10\\
\emph{Linear term:} Each activity $e=(a_i,b_j,s_k)$ will get an additional weight according to the preferences of the students and mentors as follows, that we denote by $w_e^p$. Let $w_e^p=(6-rank_i(s_k))+(6-rank_j(s_k))+3*agepref_i^j$, where each of the first two terms gives a value between 1 to 5, depending on how preferable this subject is for the student/mentor, and the last term gives 3 if the age of the student is preferred by the mentor (among years 1-4, 5-8, or 9-13). 

\item When forming groups, we shall preferably have students from the same class, with the same type of equipment, have their former grades as close as possible, and have the number of scheduled hours to be close to what the assigned students requested in that subject.\\
\emph{Rational:} Two students from the very same class are favorable to be put into the same mentoring group, as they receive the same distance education from their home school. Forming groups for students with similar strength can improve the efficiency of mentoring. Finally, the requested hours by the students in a group should be close to the scheduled hours. Regarding the equipment, if one student in a group has a laptop and another only a tablet or smartphone then the possible interactions can be limited between them and the mentor (we consider tablets and smartphones to be equally useful). \\
\emph{Relative importance in the application:} 15\\
\emph{Linear term:} 
For any two students $a_i$ and $a_{i'}$ who are both willing to accept mentoring let us introduce a binary variable $z_{i,i'}^p$ for every potential group $p=(b_j,s_k,t)$, where both $a_i$ and $a_{i'}$ could belong to, i.e., if there exist $e=(a_i,b_j,s_k)$ and $e'=(a_{i'},b_j,s_k)$ The indicator variable $z_{i,i'}^p=1$ if both $a_i$ and $a_{i'}$ are assigned to $p$ in the solution. This can be achieved with the following new constraints.

\begin{equation}\label{samegroup1} 
z_{i,i'}^p \geq y_e^t + y_{e'}^t -1 \mbox{ for every } p=(b_j,s_k,t), e=(a_i,b_j,s_k), e'=(a_{i'},b_j,s_k)
\end{equation}

\begin{equation}\label{samegroup2} 
z_{i,i'}^p \leq y_e^t \mbox{ for every } p=(b_j,s_k,t), e=(a_i,b_j,s_k)
\end{equation}

\begin{equation}\label{samegroup3} 
z_{i,i'}^p \leq y_{e'}^t \mbox{ for every } p=(b_j,s_k,t), e'=(a_{i'},b_j,s_k)
\end{equation}

Furthermore let $z_{i,i'}^k=\sum_{p: s_k\in p} z_{i,i'}^p$, where $z_{i,i'}^k$ is the indicator variable showing whether $a_i$ and $a_{i'}$ are in the same group for subject $s_k$.

The accumulated weight of group coherence criteria is
$$ GC(z)= \sum_{a_i,a_{i'}\in A} \sum_{s_k\in S} (10\cdot sc_{i,i'}- 2\cdot de_{i,i'} - dg_{i,i'}^k- dr_{i,i'}^k)\cdot z_{i,i'}^k$$

\item Giving priority to certain students in needs (depending on the preferences of the mentors for some criteria, such as social status and student past performance).\\
\emph{Rational:} Social welfare can improve if the students in need receive the mentors. However, the criteria of being socially disadvantaged or being a weak student can be controversial for some mentors, so we allow them to express their willingness to get paired with such students.\\
\emph{Relative importance in the application:} 20\\
\emph{Implementation with weights:} 

The combined social priority weight of $e=(a_i,b_j,s_k)$, denoted by $w_e^s$, is as follows:

$$w_e^s=3\cdot SD_i\cdot DM_j + WS_i\cdot SM_j+(1.5 -|gr(a_i,s_k)-WM_j|)\cdot PM_j +2\cdot CY_i+(NH_i+help_i)$$

\item When the same mentor-student pair can be involved in paired mentoring activities in multiple subjects then this is preferable.\\
\emph{Rational:} It can be useful if not too many different mentors supervise the same student.\\
\emph{Relative importance in the application:} 5\\
\emph{Linear term:} We introduce a new binary variable $m_i^j$ to denote whether $a_i$ is mentored by $b_j$ in any subject in a pair with the following constraints.

\begin{equation}\label{mentored}
m_i^j\leq \sum_{e: (a_i,b_j)\in e} y_e \leq 5\cdot m_i^j
\end{equation}

We shall minimise the number of mentoring pairs with a weight $w^m$, that is let 

$$MP(y)=w^m\sum_{i,j}m_i^j$$
 where we set $w^m=5$ in our application.

\end{enumerate}

\subsection*{Final objective function}

For $w_e=w_e^w+w_e^p+w_e^s$ (and $w_e^g=w_e\cdot w^g)$, the final objective function is:

\begin{equation}\label{objective}
\sum_{e\in E} w_e\cdot x_e + \sum_t w_e^g\cdot x_e^t + GC(z)-MP(y)
\end{equation}

To summarise, the main objective is to maximise the volume of the mentoring activities with considering the preferences and the social priorities of the students, improving the group cohesion, and decreasing the number of mentors per students.

\section{Simulations: data generation}

During the first operating time of the allocation scheme (1 May to 15 June 2020), the number of students who registered at the webpage was 14, while the number of mentors was 56. Because of the low number of student's registration, we could not observe the true potential of the model and application. Therefore we decided to conduct computational experiments on partly generated data. 

Because of the low amount of observations, the main goal of our simulation was to test our MILP model, rather than to give accurate prediction on the performance of the scheme. However, we still tried to use the data available to get as realistic generated data as possible.

%For the generation of instances, we used the available data%, even though the representativeness of the data is quite low. 
We computed the correlations amongst the mentor variables, but, we only found weak correlations. %Because of the low number of observations, the correlation may be some kind of error. 
Hence we generated the variables independently, by rolling a biased dice independently for each variable and observation (student and mentor). We used the available data to estimate these biases. In this section, we describe the way we generated each variable.

\subsection{Students}
\begin{itemize}
\item \textbf{School-ratio:} There were 14 students registered, out from 9 different schools. Therefore in the simulator, we considered the student-school ratio to at most 67\%. So when we considered 100 students, we generated 67 schools, and for each student, we selected a school with replacement. 

\item \textbf{Number of subjects:} The average number of subjects requested per student was 2, with four as maximum. We used the following distribution for the amount of subjects: (1: 50 \%; 2: 30 \%; 3: 10\%; 4: 10\%).

\item \textbf{Time ($q(a_i; s_k)$):} The average time required by the students for mentoring per week in a subject was around 2 hours with a minimum of 1 and maximum of 4. We used the distribution of (1: 33\%; 2: 33\%; 3: 25\%; 4: 9\%).

\item \textbf{Grades:} For the distribution of grades, we got almost every possibility from the data, except grade 1, which means Failed in Hungary. In the simulation, we generated all types of grades uniformly with the addition of 0, where 0 means the student left the bracket blank (which was 25\% of the real cases).

\item \textbf{Group:} 9 out of 14 students selected the possibility of getting mentoring in groups. Hence we set $\frac{2}{3}$ for the probability of a student accepting group-mentoring.
   
 \item \textbf{Help:} For the three possible values we received 5-5-4 responses, respectively. Therefore, we decided to generate these values randomly with equal probability. 
    
\item \textbf{Equipment:} All of the participants chose 0. We left this parameter out in the simulation because it is not crucial for the optimisation model.
    
\item \textbf{Year:} Most of the registered students were from years 4-8 (92\%), only one student was in year 11. Since the programme started at the end of the spring semester and after the matriculation exam, we assume that the secondary school students were underrepresented in the application. Therefore we generated the students' years uniformly from year 4 to 12.
    
\item \textbf{Prior1 ($SD_i$):}
13 students out of 14 chose option 0, and the last remaining student chose option 1. Therefore we generated higher ration of underprivileged students by using the  (0: 65\%; 1: 20\%; 2: 10\%; 3: 5\%) distribution for this variable.

\item \textbf{Prior2 ($NH_i$):}
 In the generator, we used the result of a 2016 census of the Hungarian Central Statistical Office (\url{http://www.ksh.hu/thm/2/indi2_1_4.html?lang=hu}) to generate the data. According to this census the ratio of the single-parent family is $18.3\%$. 
For the number of children, we generated the data according to the table \ref{tab: nbr_children} (for the \textit{4 or more} we considered simply 4).

    \begin{table}[h]
    \centering
\begin{tabular}{l|r}
Number of children & probability \\ \hline
1                  & 69.2\%      \\
2                  & 23.9\%      \\
3                  & 5.2\%       \\
4 or more          & 1.7\%      
\end{tabular}
\caption{Families by number of children, (\url{http://www.ksh.hu/thm/2/indi2_1_4.html?lang=hu})} \label{tab: nbr_children}
\end{table}

Therefore, for each student, we generated the number of parents and the number of siblings independently, according to the above described statistics. 
    
\item \textbf{Prior3 ($WS_i$):}
In the real data, $64\%$ of the students had 0 points, $28\%$ had two and only one student had 3 points (no student got 1 point for this variable).
 We approximated the values of this variable with the Poisson distribution, where we set the expected value equal to the mean of the real data ($0.786$).  
 
\item \textbf{Prior4 ($CY_i$):}
This point depends directly on the year of the student, so it is computed accordingly.

\item \textbf{Matriculation:} This depends on the year of the student. Therefore we only considered this variable when the student was from year 11 or 12. Most students officially take the matriculation exam in year 12 in Hungary, however,  there is an option to advance some exams to previous years. Hence we randomised this value for the students from year 11, by assuming that 40\% of them do not want to practice for the matriculation exams (so they got the "N"-letter, meaning no matriculation), 40\% going for the base exam and 20\% of them selecting advanced exam.
\end{itemize}

\subsection{Mentors}

When we generated the characteristics of the mentors, we used the distributions taken from the real data. Since 56 mentors registered into the program, we have not made as many assumptions as in the case of the students.

\begin{itemize}
    \item \textbf{Group:} 54\% of the mentors agreed to the possibility of mentoring in groups.
    
    \item \textbf{Time ($Q(b_j)$):} For the time capacity of the mentors, first we generated ranges with a distribution, and then we choose uniformly the exact value from the range selected. The time-range of a mentor was 1-3 hours with 40\% probability, 4-6 hours with 40\% probability, and 7-10 hours with 20\% probability.

    \item \textbf{Social ($DM_j$):}
    We used the (0: 50\%; 1: 40\%; 3: 10\%) distribution, reflecting the real data.
    
    \item \textbf{Weak ($GPM_j$):}
    Here 85\% of the mentors chose option \textit{N}, therefore we also used 85\% for generating option \textit{N} and 5\% for each the other three options.
    
    \item \textbf{Student-Age ($YM_j$):}
    We used the (0: 5\%; 1: 20\%; 2: 15\%; N: 60\% (no preference given)) distribution, reflecting the real data. 
    
\end{itemize}

\subsection{Subjects}

Overall, there were 15 different subjects offered for selection. We generated the distribution of the requested subjects according to the distribution of the subjects offered by the mentors. The assumption behind this is that the demands and the supply of the subjects shall be balanced in the long run.  However, we also added a random noise to modify the distribution for every instance. %\PB{Pontosan hogyan? Ha ez nem lenyeges, akkor lehet, hogy erdemesebb inkabb kihagyni a kommentet, hogy hasznaltunk parametert.} \Marci{Lényegében a mentorok tárgyainak eloszlásához hozzáadtam egy random paramétert - amivel először legenerálom a diák-tárgyak eloszlását. Ezzel kapunk egy nagyjából hasonló eloszlást, amit a végső genráláshoz használtunk. Azért csináltam így, mert így mindegyik generálásnál kicsit másmilyen lesz a Kereslet-kínálat viszont az eltérés nem lesz nagy (pl. ha a matekot sok mentor kínálja, akkor a diák-keresletnél is a magasabb keresletűek között marad). -talán nem olyan fontos, de megemlíteném, hogy a mentor-tárgy-eloszlásból állítjuk elő a diák-tárgy eloszlást, randomizálással.}

Not all of the subjects are for each year. Therefore the distributions are normalised for each year with regard to the subjects available.
A student can request mentoring in multiple subjects. According to the data, the maximum number of subjects requested was 4 with an average of two. As we already described, we used the (1= 50\%, 2 = 30\%, 3= 10\%, 4 = 10\%) distribution to generate the number of subjects requested by each student, and then with the consideration of the student's year, we picked the subjects randomly with a distribution that is close to distribution of the subjects offered by the mentors. 

The mentors can also have multiple classes. In the data, the average number of subjects per mentor was 2.9. However, the maximum number of the subjects was 9 for some mentors. Therefore some extreme values have increased the average a lot. We generated the number of subjects of each mentor with respect to their total time offered. If the mentor's total time was less than 4 hours then we generated 1-3 subjects uniformly. If the total time was less than or equal to 6, then we generated 1-4 subjects  uniformly. Finally, in the case of time at least 7, we generated 1-5 subjects. 
Whenever we allocated a subject to a mentor, we assumed that the mentor is willing to teach students of all age in this subject.

\subsection{Comparison of generated and real data}

To compare the simulated data to the real data, we generated 1000 instances with the same amount of mentors (56) and students (14) as in the real data. 
Then we checked whether the values of the real data are within the interquartile range of the values of the generated data. 

Regarding the evaluation measures, only the \textit{Social}-points were not inside this range with any type of objective value. This is reasonable since in the real data, we have not received enough values (see, for example, Prior 1 and 2). Therefore we adjusted the distributions. Hence the generated data got in average higher results than in reality. %\PB{Ez nem teljesen világos, az eloszlásokat mikor és miként módosítottuk?}\Marci{pl. Prior1}

Also, when we considered the group-weight of 1 in the objective function, several other measures were also outside the interquartile range. The reason behind this instability is the few numbers of students in the real data.

\section{Simulations: single run}\label{measures}

In this section first we describe how we simplified the evaluation measures by clustering them based on one-shot simulations, and then we also present the performance analyses with regard to some basic parameters.

\subsection{Generation of instances}

For the basic setting we generated 100 large instances with 80 students and 40 mentors in each, and we conducted a single match run for each of them. This setup differs from the real data, since the student/mentor ratio is much higher in the generated data. Our aim was to analyse the effect of using different objective functions when the programme has an ideal student/mentor ratio. 

Then we adjusted the main parameters to create alternative solutions, as follows.

\subsubsection*{Alternative parameters}

\begin{itemize}
    \item we modify $w^g$ from 0.7 to 0.5, 0.6, 0.8, 0.9, 1
    \item for $w^g=0.7$ we modified the preference based weights from the default setting of a) $w_e^p=(6-rank_i(s_k))+(6-rank_j(s_k))+3*agepref_i^j$\\ to b) $w_e^p=(6-rank_i(s_k))^2+(6-rank_j(s_k))+3*agepref_i^j$,\\ and to c) $w_e^p=(6-rank_i(s_k))^2+(6-rank_j(s_k))^2+3*agepref_i^j$%Part of the matching scheme it was agreed that effectiveness of the paired and group mentoring will be compared, an so a randomised experiment has been proposed. In this experiment some students, who have the option for both pair and group mentoring in a subject will be allocated randomly, and the effect of this random choice will be measured with the long term success of mentoring.
    
\end{itemize}

Thus we generated five times 100 instances by varying $w^g$ and another two times 100 instances by changing the preference based weights. So altogether we considered 800 instances. The run time for solving the MILP model for these instances was relatively short, the average run time was 167 seconds with a maximum of 3 hours and 12 minutes. 

\subsection{Clustering the evaluation measures}

Besides the seven objectives given in Section~\ref{sec:LP}, we considered the following five performance measures in the evaluation of solutions.

\subsubsection*{Evaluation measures}
\begin{enumerate}
    \item Number of paired mentoring hours and group mentoring hours separately.
    \item Number of pairs and number of groups separately.
    \item Total capacity of the mentors used.
\end{enumerate}
 
\begin{figure}[h!]
    \centering
    \captionsetup{justification=centering, margin = 0.4cm}
	\includegraphics[width = 1.0\columnwidth]{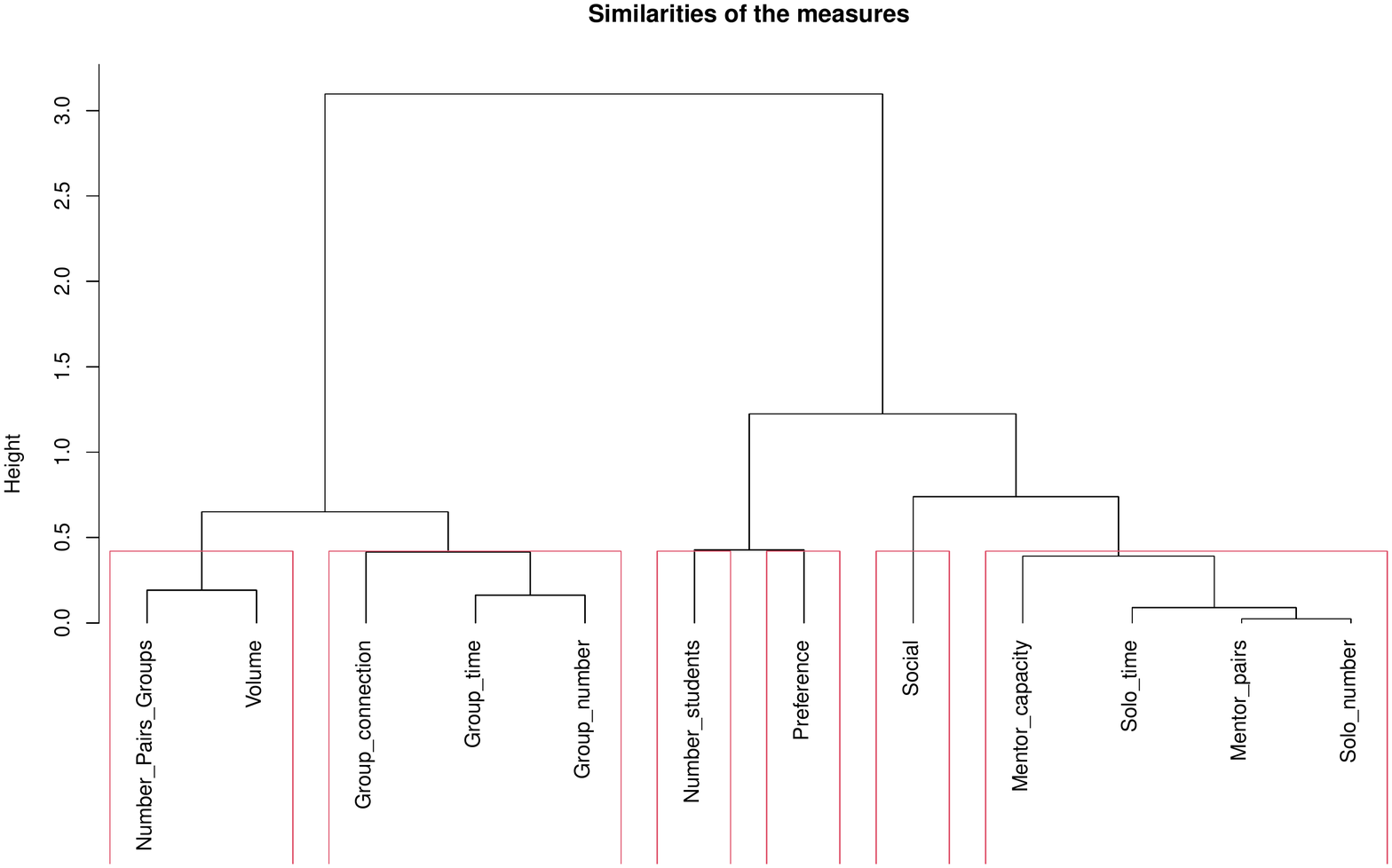}
    \caption{The dendrogram of the measures. The measures are described in section \ref{sec:LP}\\ 
    \texttt{Number\_students}: Objective 1
    \texttt{Number\_Pairs\_Groups}: Objective 2; 
    \texttt{Volume}: Objective 3;
    \texttt{Preference}: Objective 4;
    \texttt{Group\_connection}: Objective 5; 
    \texttt{Social}: Objective 6;   
    \texttt{Mentor\_pairs}: Objective 7; 
    \texttt{Solo\_time/Group\_time}: Evaluation measure 1;
    \texttt{Solo\_number/Group\_number}: Evaluation measure 2;  
    \texttt{Mentor\_capacity}: Evaluation measure 3;  
     }
    \label{fig:dendrogram}
\end{figure}

We analysed how the above described five evaluation measures behave together with the seven different objective functions, that we presented earlier. 
However, for the evaluation of the different objective functions, we decided to reduce the dimension of the twelve measures. So, we calculated the similarities between the various measures with hierarchical clustering. All of the results were considered in the estimation. Figure \ref{fig:dendrogram} presents the dendrogram of the measures.

We decided to reduce the dimension of the 12 measures to 6 (as the red rectangles show in Figure \ref{fig:dendrogram}). We chose the six dimensions, because according to Principal component analysis, with 6 component 96\% of the total information can be saved. Additional component only increased the saved information with less than 2\%.

According to the hierarchical clustering, we evaluated the  \texttt{Number\_students},  \texttt{Preference} and \texttt{Social} measures independently from the other measures.

From the  \texttt{Mentor\_capacity}, \texttt{Solo\_time}, \texttt{Mentor\_pairs} and \texttt{Solo\_number} measures with using Factor analysis method, we created a factor variable. We covered 84\% of the total information in this factor. We named this factor as \emph{solo-factor} because it is related to the measures of the non-group classes. For every measure a higher value increase the accumulated factor score. Interestingly the number of Mentor-student pairs (Objective 7) has a behavior very similar to the solo-class time and number. %The reason behind may be because of the short time capacity of the classes and the high-ratio of solo-groups.

We created another factor for the measures of \texttt{Group\_connection},  \texttt{Group\_time}, and the  \texttt{Group\_number}. We named this factor as \emph{Group-factor}. We could cover 73\% of the total information inside one factor. Higher \texttt{Group\_connection} value decreases the Group-factor's score, while for the other two measures higher value will increase it.

Finally, from the \texttt{Number\_Pairs\_Groups} and \texttt{Volume} we created the so-called \emph{Quantity-factor}. For all of these measures higher values increase the factor's score. We could cover 81\% information within this factor.

\begin{figure}[h!]
	\centering
	\captionsetup{justification=centering, margin = 2cm}
	\begin{minipage}{.4\textwidth}
		\includegraphics[width=1.2\textwidth]{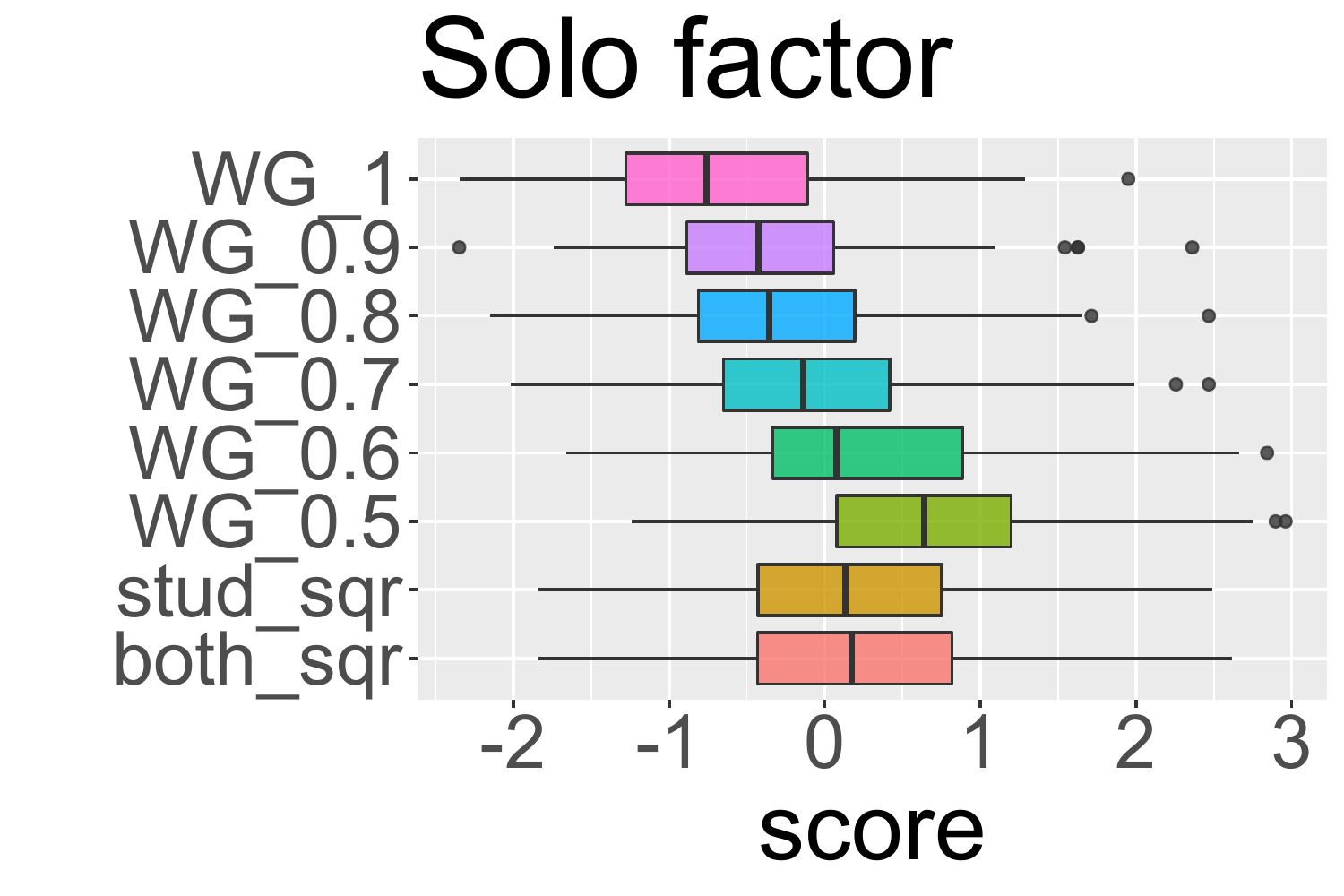}
	\end{minipage}
	\hspace{1cm}
	\begin{minipage}{.4\textwidth}
		\includegraphics[width=1.2\textwidth]{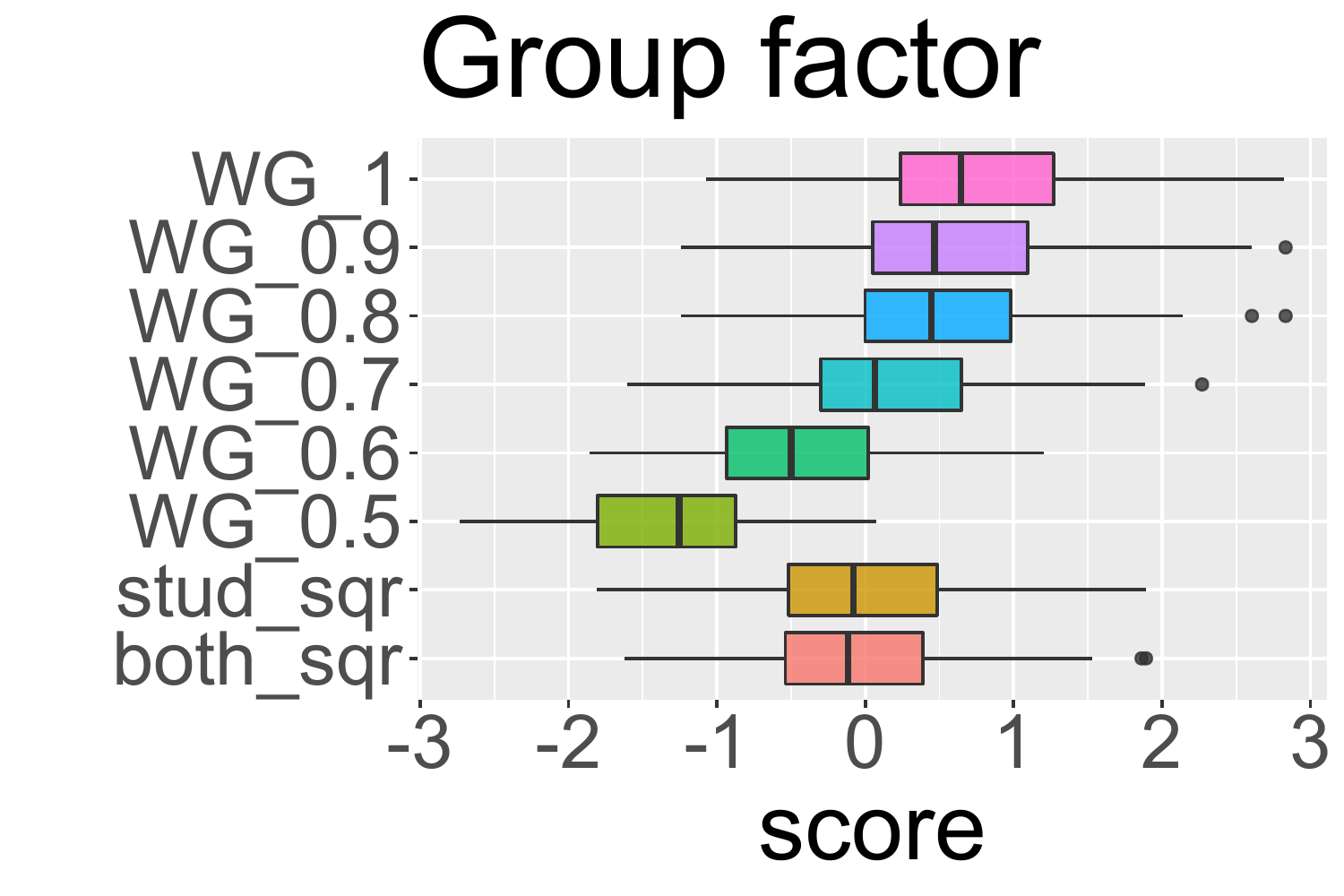}
	\end{minipage}
	%\vspace{0pt}
	\begin{minipage}{.4\textwidth}
		\includegraphics[width=1.2\textwidth]{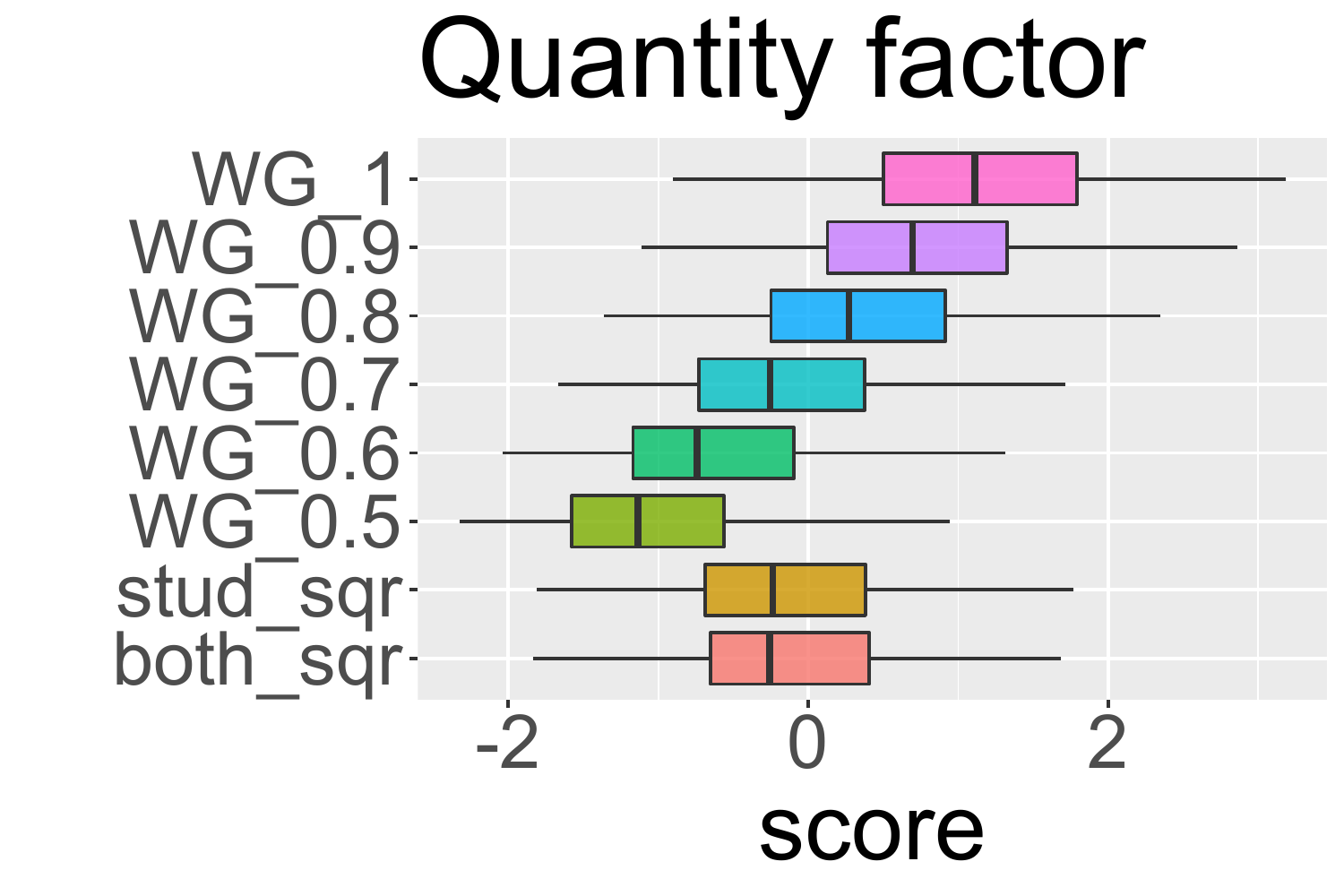}
	\end{minipage}
	\hspace{1cm}
	\begin{minipage}{.4\textwidth}
		\includegraphics[width=1.2\textwidth]{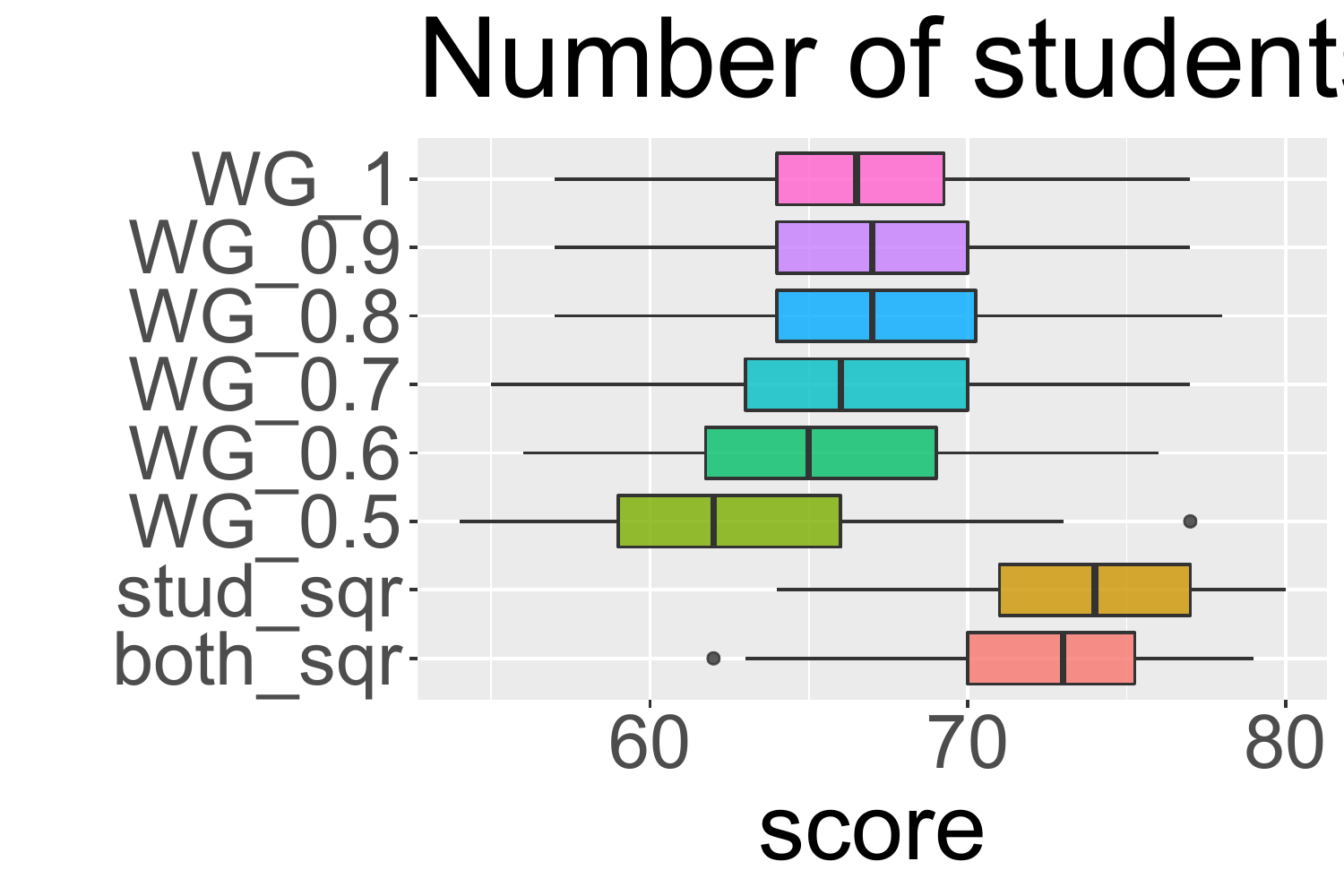}
	\end{minipage}
		\begin{minipage}{.4\textwidth}
		\includegraphics[width=1.2\textwidth]{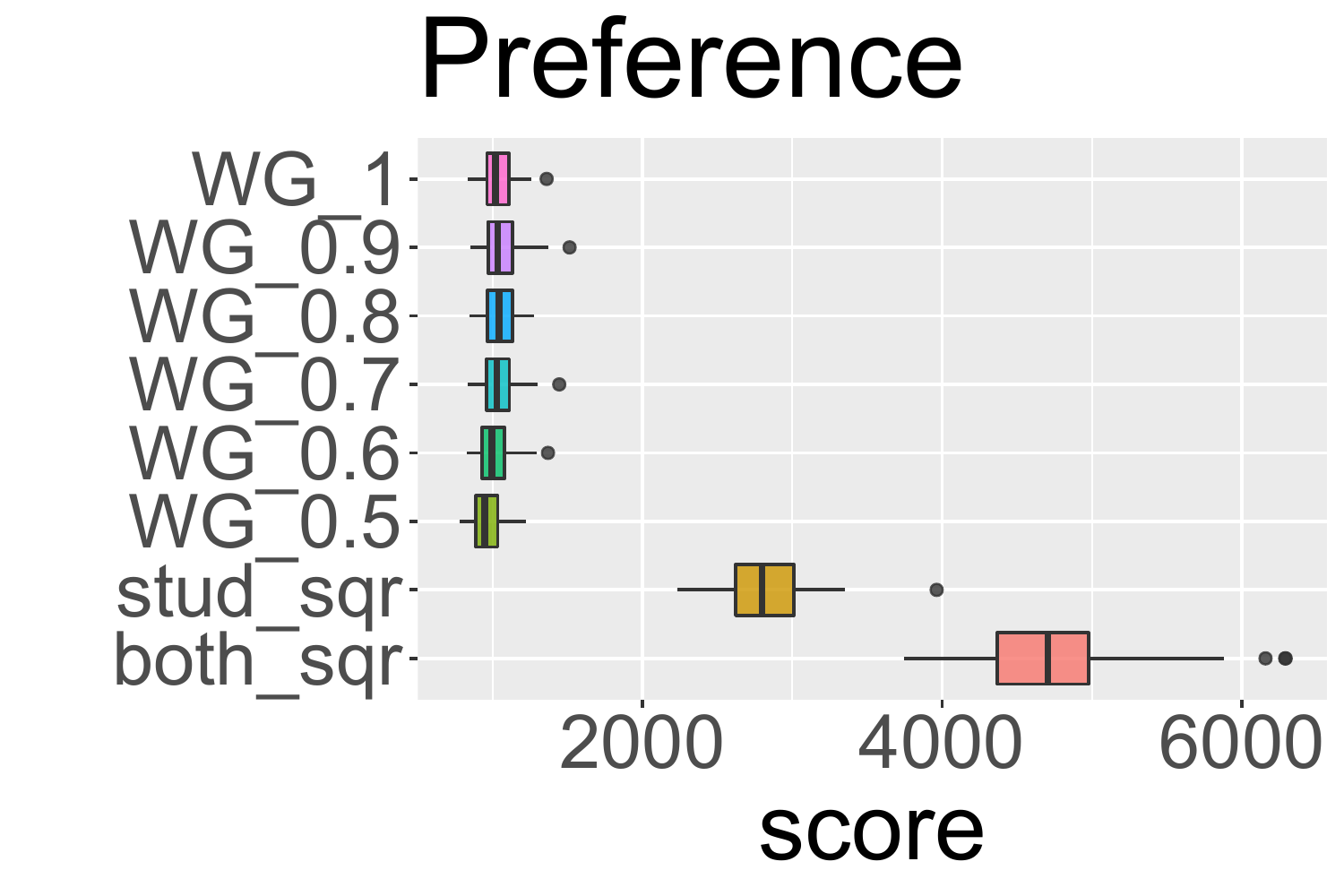}
	\end{minipage}
	\hspace{1cm}
	\begin{minipage}{.4\textwidth}
		\includegraphics[width=1.2\textwidth]{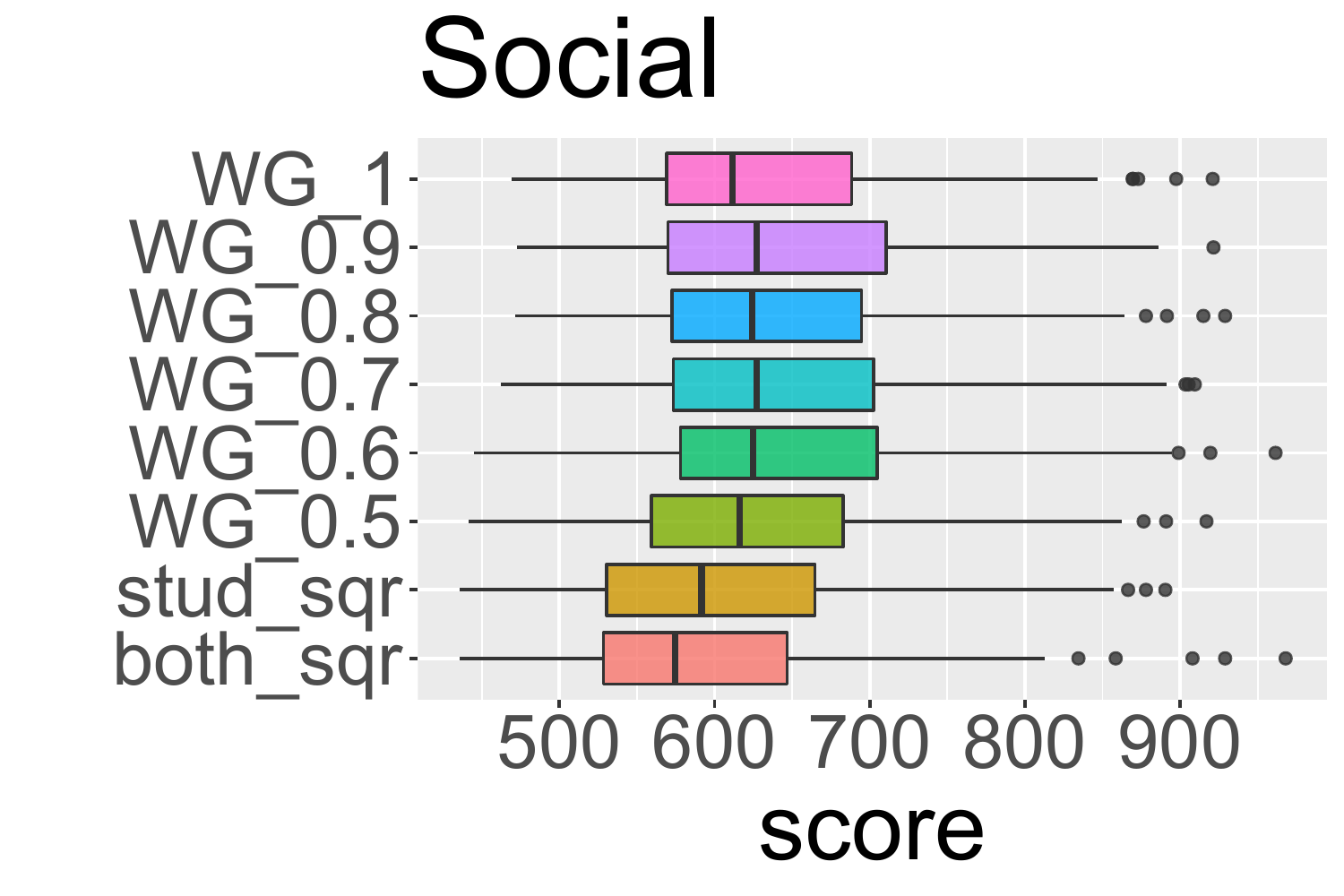}
	\end{minipage}
	\caption{The distribution of the 3 factors and the Number of students, Preference and the Social measures.} 
	\label{fig:boxplots_measures}
\end{figure}

\subsection{Analyses of one-shot simulations}

Figure \ref{fig:boxplots_measures} presents the distributions of the three factors as well as the Number of students, Social and Preference scores of the different objective functions.

On the graphs of the Solo-factor and Group-factor, the influence of group-weight ($w^g=WG$) is clearly visible.  As the group-weight increases, the Solo-score decreases and the Group-score increases.
The two quadratic-preference models had very similar results in these cases compared to the original \texttt{WG\_0.7} model, although there was a small decrease in the group-score.

The Quantity factor's score increases with the group-weight.  Therefore if the number of pairs and groups, or the volume is more important in the optimisation, then the usage of higher group-weights seems to be more appropriate.

Group-weight also increases a bit the number of matched students. However, the best solution for this measure is when the preference of the students was considered quadratically.

Naturally the Preference measure is very high for the two quadratic preference cases. However, it seems that the group-weight parameter did not have any effect on this measure.

Social points appear to be very stable, the group-weight does not have much effect on the social aspect of the programme. However, the quadratic preference cases have a slightly worse result for this measure.

\section{Simulations: dynamic allocation}\label{dynamic}

We investigated also how the frequency of the matching runs affects the results in a dynamic setting. We considered the first 300 days of the programme, by generating a registration date for each student and mentor. We assumed that both the students and mentors joined the application uniformly random in the period considered. We also generated a leaving day for each applicant as follows. Presumably, if a student does not get any mentor within a reasonable time in this programme, then he or she may well seek mentors using other channels. We assumed that the students stay in this programme for an average of two weeks, with two days of standard deviation with the assumption of the normal distribution (but they remain at least seven days). The mentors may also leave a programme if they do not get any student soon enough after registration, however, we assumed that they are be more patient than the students. Hence they leave the programme 3 weeks after registration in average (with a minimum of two weeks). Again we used normal distribution with setting two days for the standard deviation.

\begin{figure}[h!]
	\centering
	\captionsetup{justification=centering, margin = 2cm}
	\includegraphics[width=1.0\textwidth]{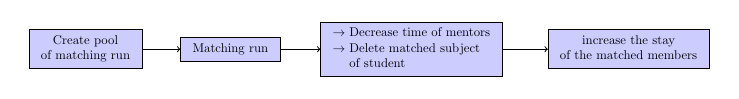}
	\caption{Flow of a matching run}
	\label{fig:flow}
\end{figure}

We also extended the stay of those students and mentors, who got matched. For any student, who was in a solution of a matching run, we added seven more days to their leaving time. We extended the staying time of the matched mentors with 14 days. Figure \ref{fig:flow} presents the flow of a matching run.

First, we set up the pool of students and mentors of a matching run. A pool of students consists of those students who have already registered into the programme, but have not left it yet and have demand for some subject. The pool of the mentors are those who have time for teaching, and already registered but have not left the pool yet.

Then we calculate the optimal matching according to the policy setting and we remove the satisfied subject from the list of the demanded subjects of each selected student. For the allocated mentors, we decrease their total time offered for teaching with the allocated time.

Then we increase the remaining time of stay of the allocated students and mentors, with 7 and 14 days, respectively.

We considered four different frequencies for the matching-runs: 1, 2, 7 and 14 days. We evaluated 100 generated instances according to these four frequencies. In the optimisation model, we considered the Final objective function (\ref{objective}) with the default group-weight $w^g = 0.7$.

Since we had no access to the registration dates and duration of stays of the students and mentors, we generated two different set of instances with regard to the frequency of the arrivals. We considered instances where 1 and 4 were the average number of students joining the programme per day. Therefore, for an instance of type 1, we generated 300 students, and for an instance of type 4, we generated 1200 students for the period of 300 days. The students register into the programme uniformly at random, hence 1 and 4 students register daily on average, respectively.

In each case, we assumed that the mentors arrive half as frequently as the students. Hence in instances of type 1, one mentor arrives in every two days, in instance of type 4, two mentors are expected to join the programme every day in average.

\begin{figure}[h!]
	\centering
	\captionsetup{justification=centering, margin = 2cm}
	\begin{minipage}{.4\textwidth}
		\includegraphics[width=1.2\textwidth]{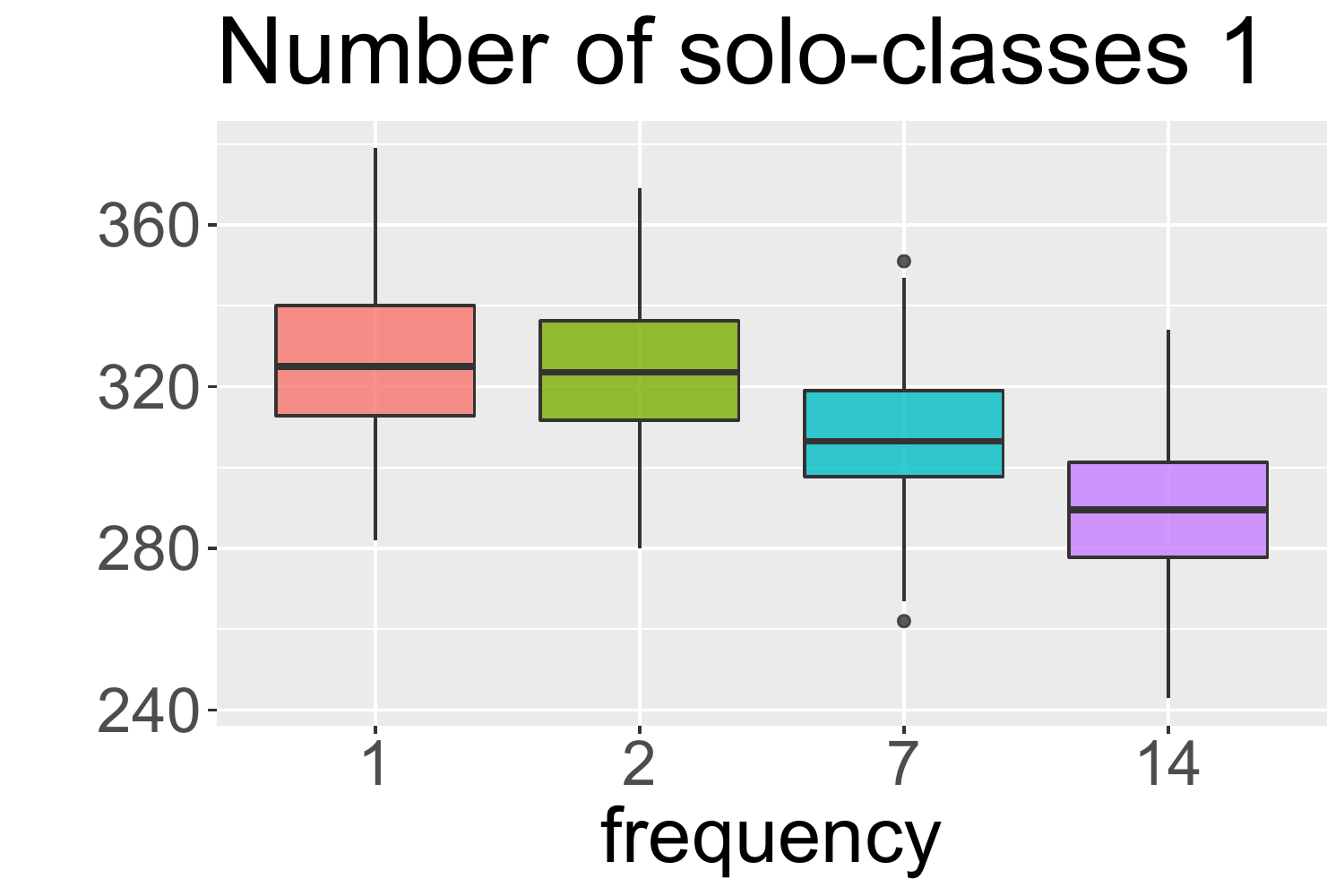}
	\end{minipage}
	\hspace{1cm}
	\begin{minipage}{.4\textwidth}
		\includegraphics[width=1.2\textwidth]{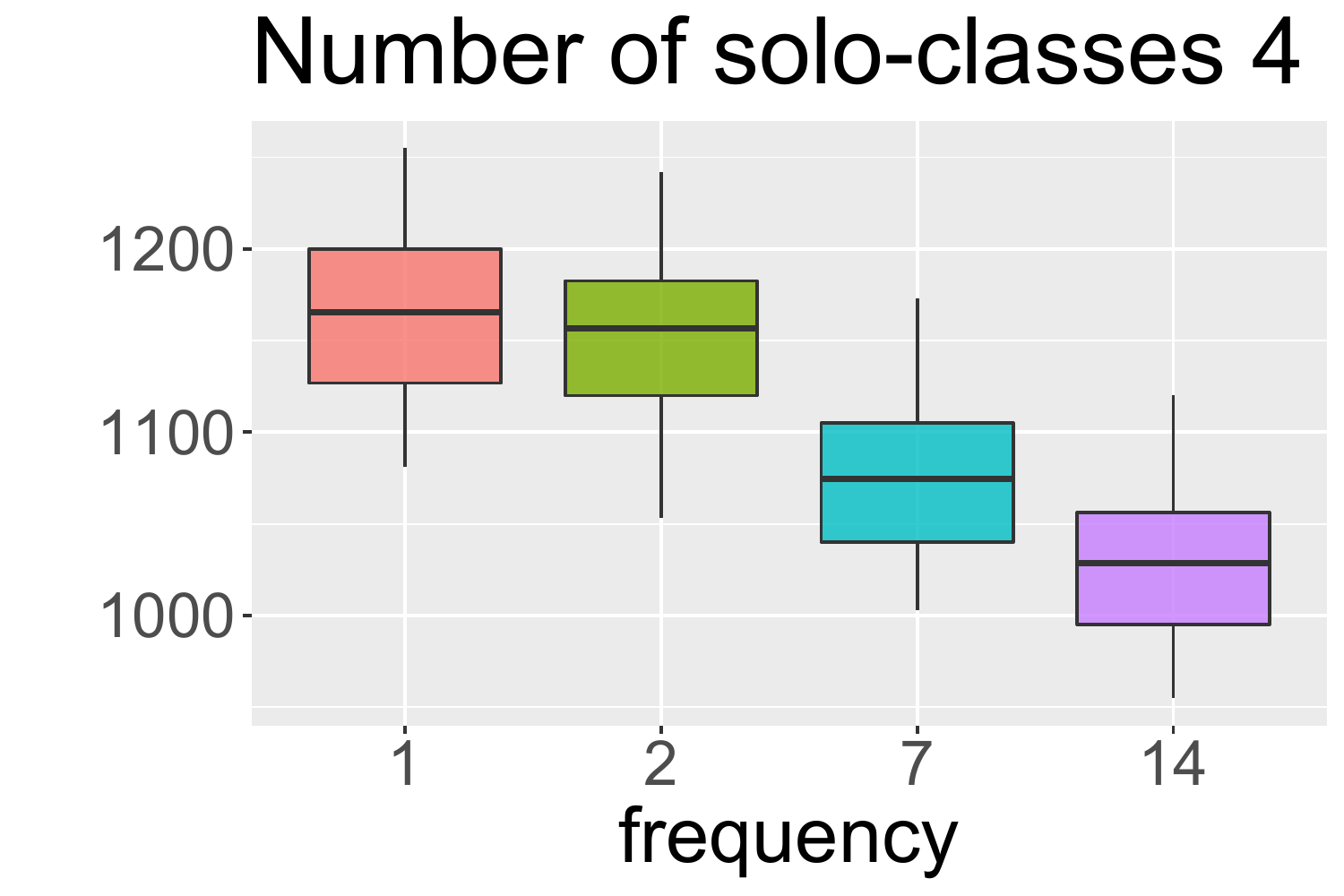}
	\end{minipage}
	\caption{The formed solo-groups} 
	\label{fig:freq_solo}
\end{figure}

Figure \ref{fig:freq_solo} presents the solo-groups formed in different examples.

\begin{figure}[h!]
	\centering
	\captionsetup{justification=centering, margin = 2cm}
	\begin{minipage}{.4\textwidth}
		\includegraphics[width=1.2\textwidth]{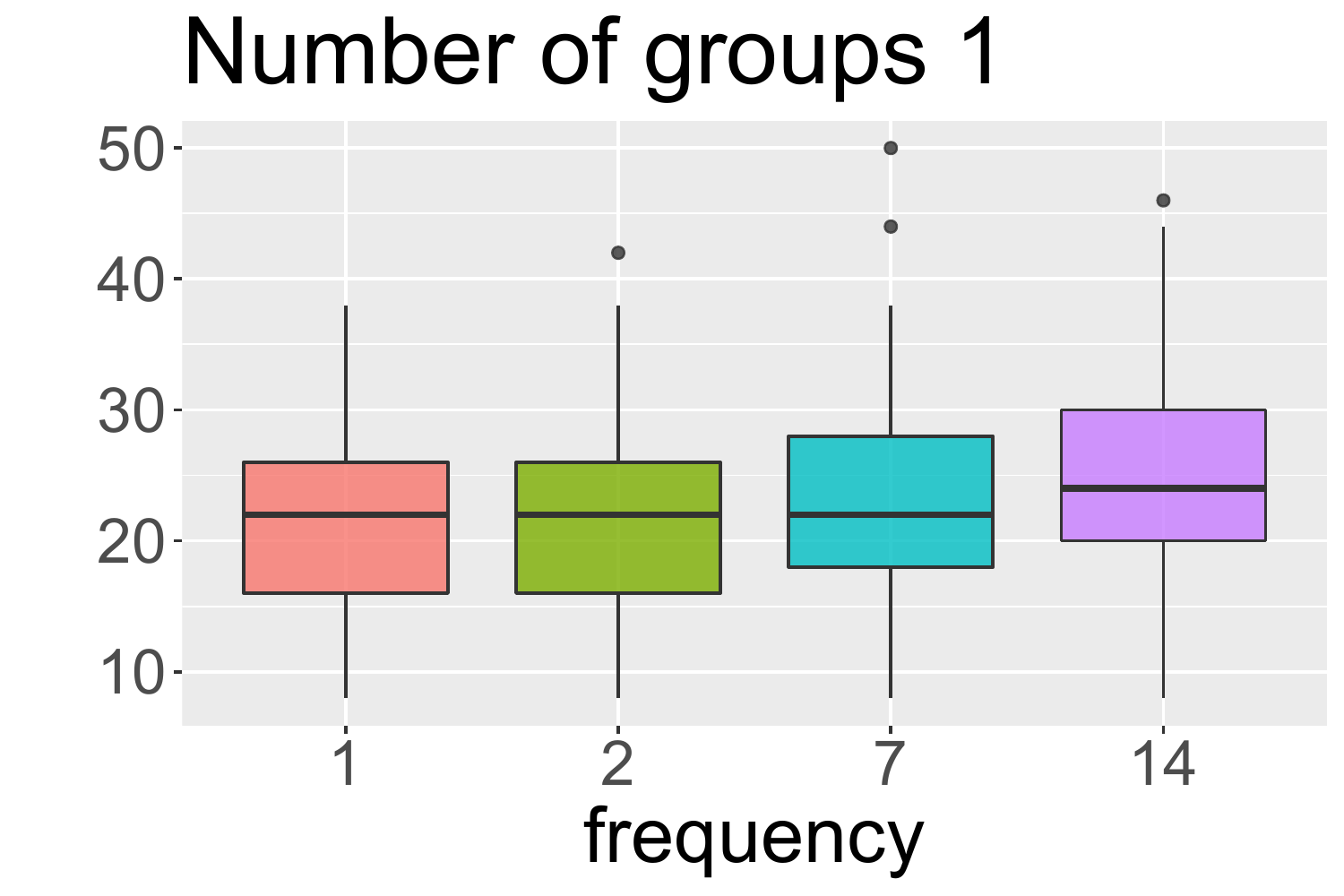}
	\end{minipage}
	\hspace{1cm}
	\begin{minipage}{.4\textwidth}
		\includegraphics[width=1.2\textwidth]{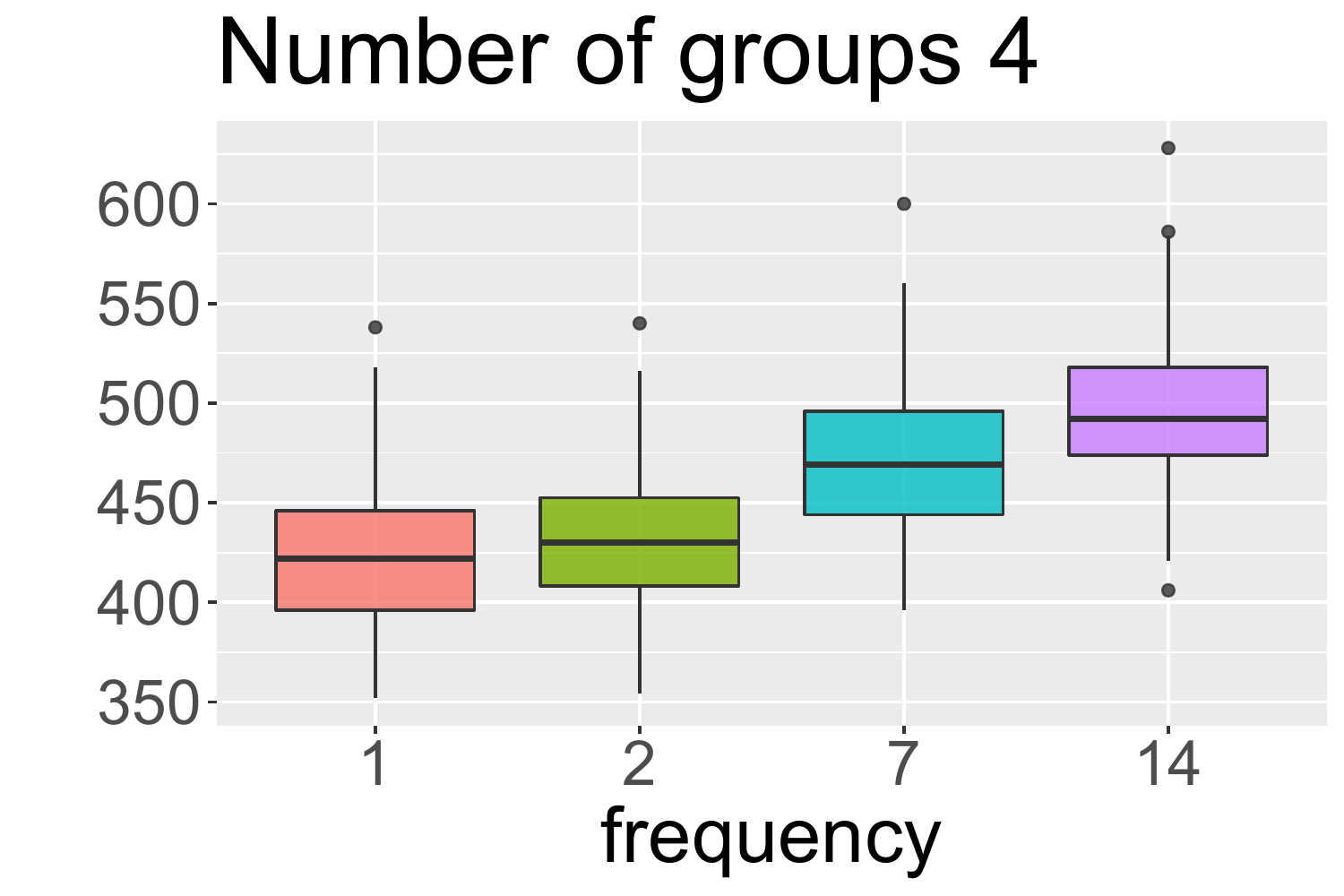}
	\end{minipage}
	\caption{The formed groups} 
	\label{fig:freq_group}
\end{figure}

In both cases, more frequent runs resulted in more solo classes. Between the 1 and 2 daily runs, the difference is not large, however as the matching runs became less frequent, the difference is visible. Figure \ref{fig:freq_group} presents the results of the number of groups formed. 

In both cases, less frequent runs resulted in more groups. Therefore independently from the sizes of the matching problems, more frequent runs were better with respect to paired-mentoring, and less frequent runs were better for forming more groups.

\begin{figure}[h!]
	\centering
	\captionsetup{justification=centering, margin = 2cm}
	\begin{minipage}{.4\textwidth}
		\includegraphics[width=1.2\textwidth]{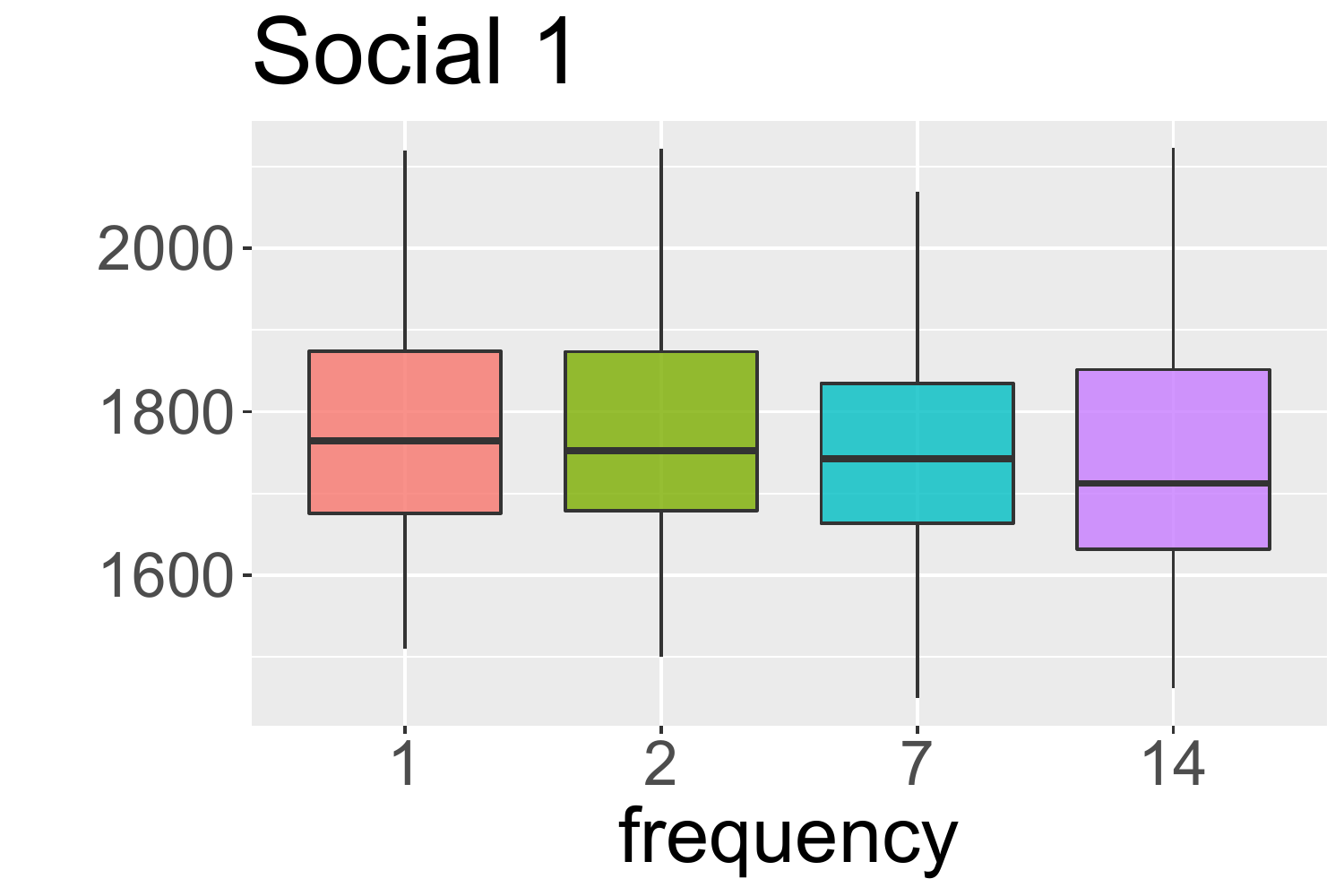}
	\end{minipage}
	\hspace{1cm}
	\begin{minipage}{.4\textwidth}
		\includegraphics[width=1.2\textwidth]{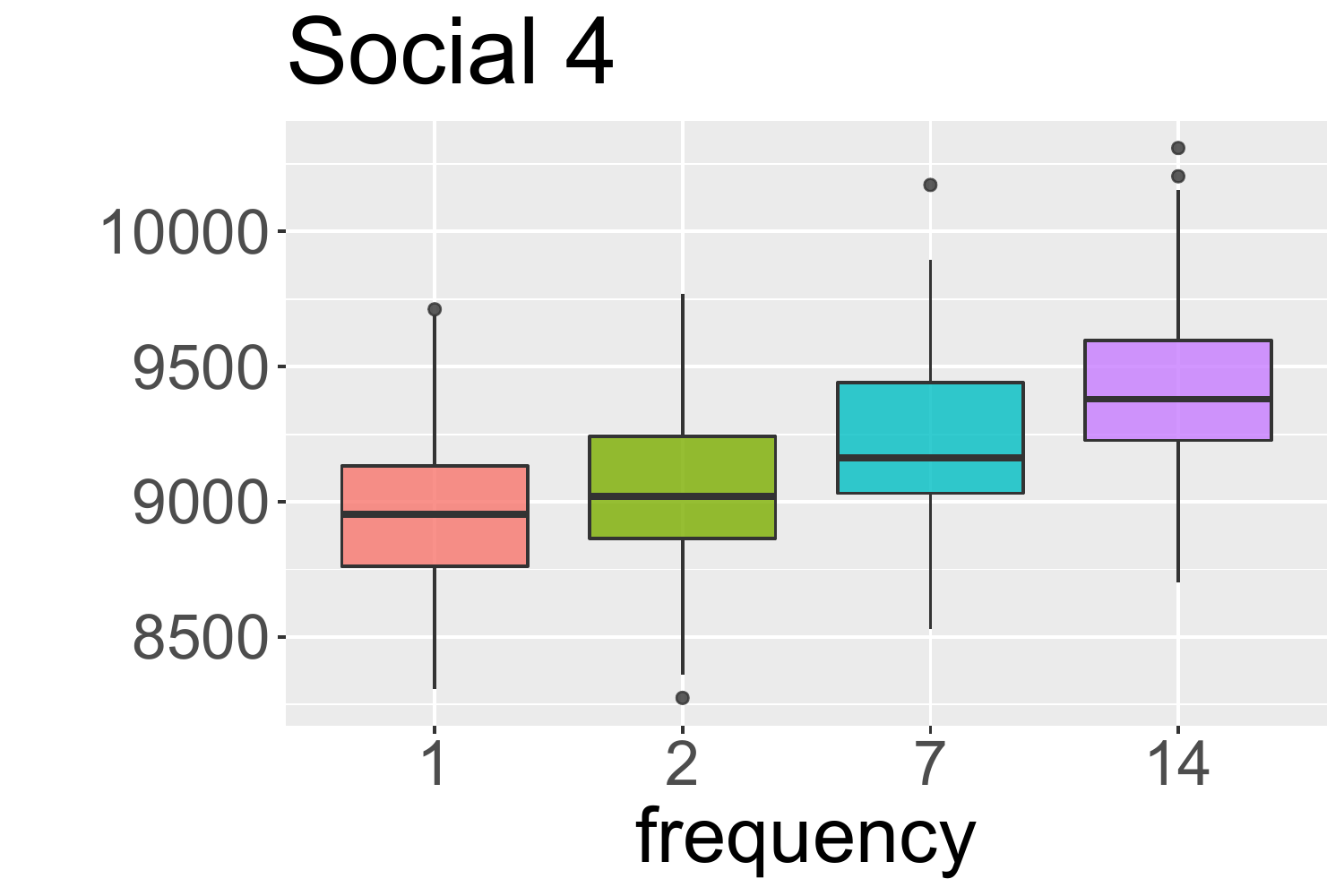}
	\end{minipage}
	\caption{The sum of social scores in each example} 
	\label{fig:freq_social}
\end{figure}

How about the qualities of the solutions of each frequency? In the previous figures, we focused on the number of pairs and groups. For the quality of matchings, first we considered the Social aspect of the programme.
Figure \ref{fig:freq_social} presents the sum of the Social points in each type of matching runs.

The Social score did not depend on the frequency of the matching runs that much when in average one student registers per day. A small decrease is noticeable as runs become less frequent in this case. 

However, when the average number of daily student registration is four, then the trend changes. The reason behind the change of trend is the higher number of groups. 

\begin{figure}[h!]
	\centering
	\captionsetup{justification=centering, margin = 2cm}
	\begin{minipage}{.4\textwidth}
		\includegraphics[width=1.2\textwidth]{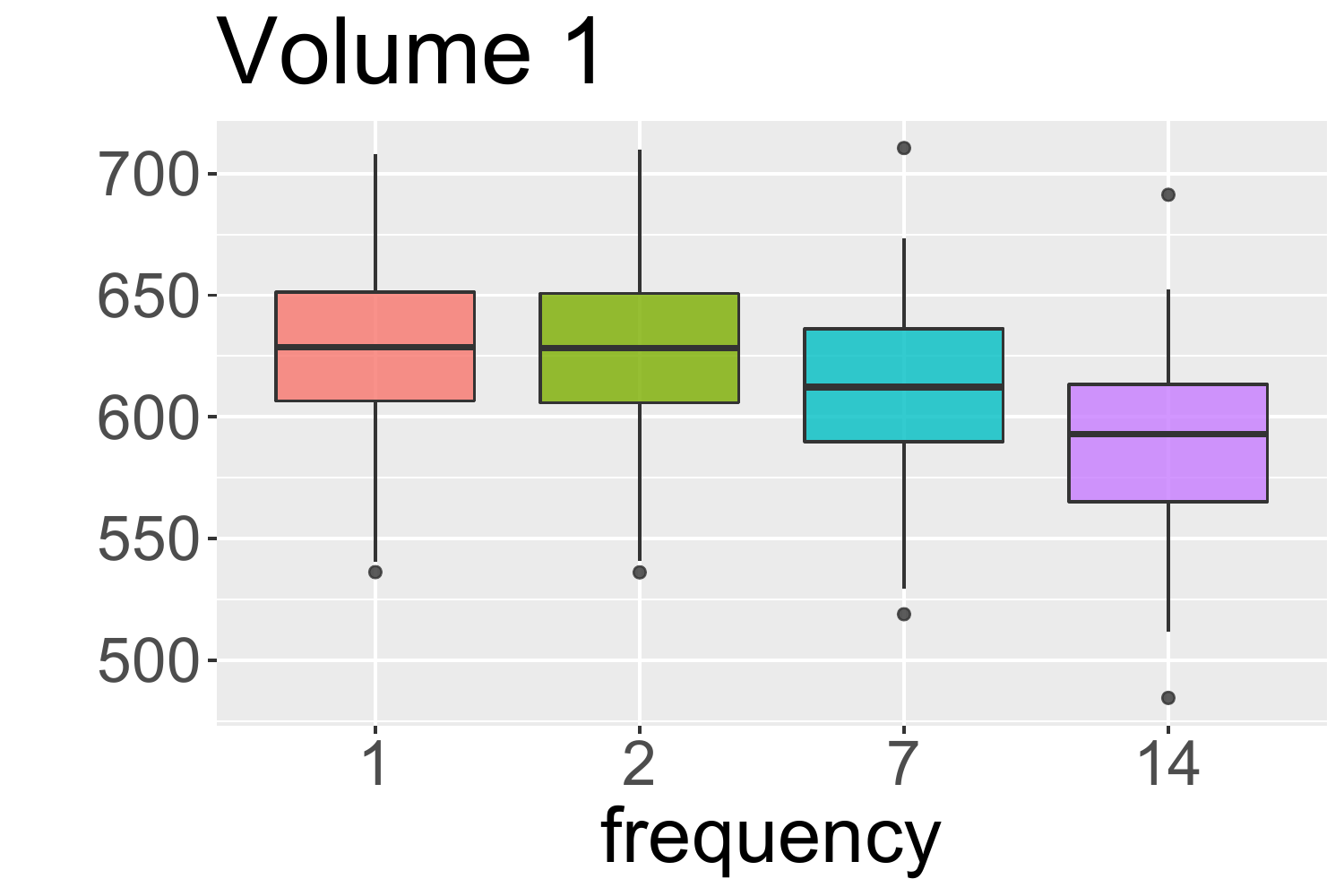}
	\end{minipage}
	\hspace{1cm}
	\begin{minipage}{.4\textwidth}
		\includegraphics[width=1.2\textwidth]{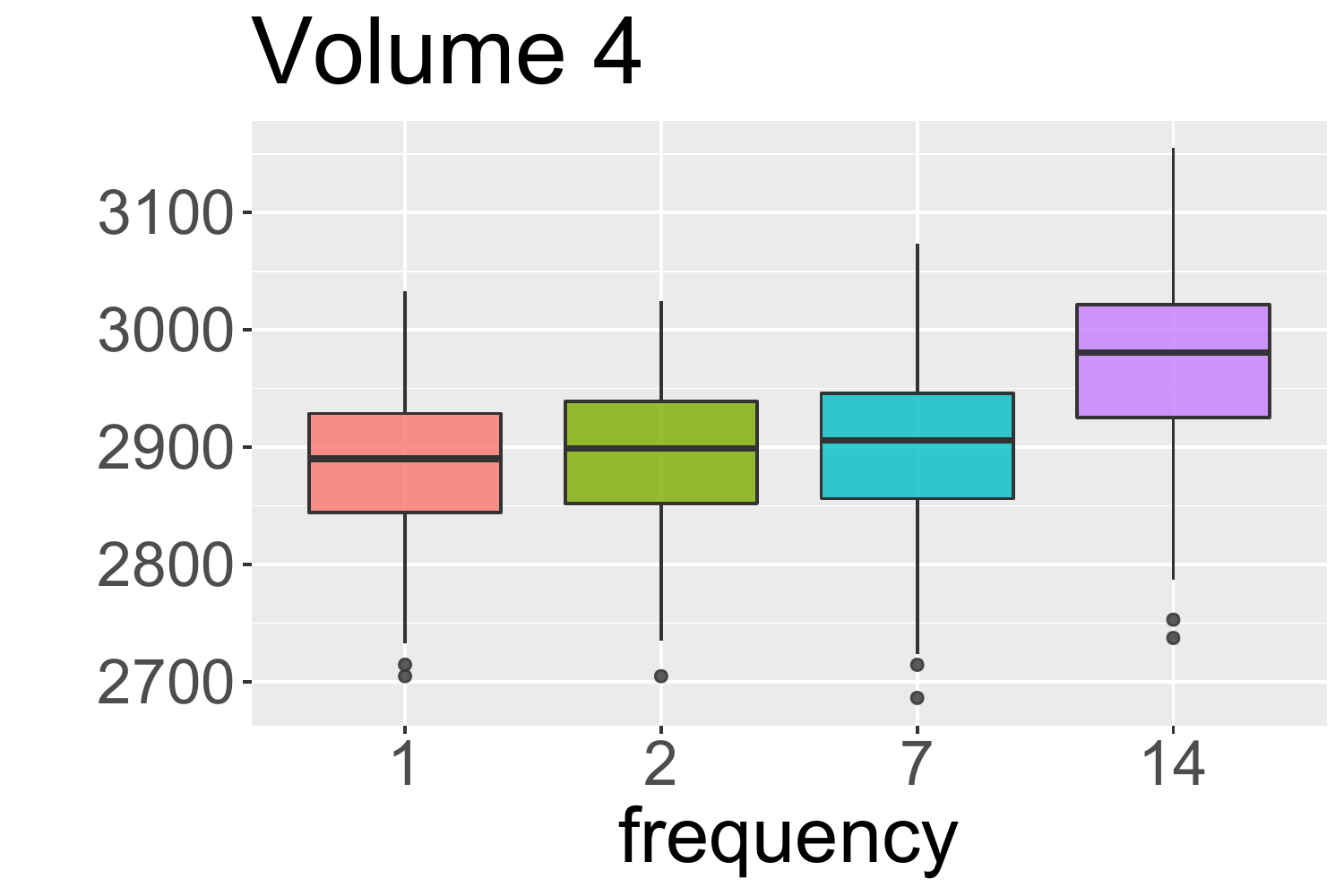}
	\end{minipage}
	\caption{Volume} 
	\label{fig:freq_volume}
\end{figure}

Figure \ref{fig:freq_volume} presents the volumes of optimal solutions. With fewer students, forming a group is more complicated. Hence more frequent runs have better social points, because of the solo-groups. With a larger pool, more groups can be formed, therefore we see an increase in the Volume, as well as in the Social score. 

\subsection{Consideration of the waiting-time}

As in other dynamic allocation systems with bounded length duration of stays (e.g., organ allocation) one may try to improve the solutions with a prioritisation based of \emph{waiting times}. %In the real-programme, the runs occurred independently. Hence there was some limitation on the registration data. The time between the registration and the matching was not specified for both mentors and students.

In the following part we investigate how the results change if we prioritise those students and mentors, who registered earlier to the programme. It may decrease the social-scores and the preference-scores of the matchings, but there may be less early quits from the programme and also less unmatched participants.

Therefore we modified the weight of the activities, to $w_e = w_e^w+w_e^p+w_e^s + w_e^t$, where  
\begin{equation}\label{eq:wait_weight}
     w_e^t = WT\bigg((t^r - t^0_i) +(t^r - t^0_j)\bigg) \quad (a_i, b_j) \in e
\end{equation}

Here $t^r$ denotes the day of the matching run, $t^0_i$ is the registration date of the student $a_i$ and $t^0_j$ is the registration date of the mentor $b_j$. We also use a weight for the days passed, that we denote by $WT$. 

To test how the prioritisation by waiting time changes the solutions, we considered a 300 days period again with an average of three students registering into the programme in every day. We set 0,1,2 and 10 for the weight of the waiting time priority regarding all the match-frequencies that we investigated earlier. 

\begin{figure}[h!]
	\centering
	\captionsetup{justification=centering, margin = 2cm}
	\begin{minipage}{.4\textwidth}
		\includegraphics[width=1.2\textwidth]{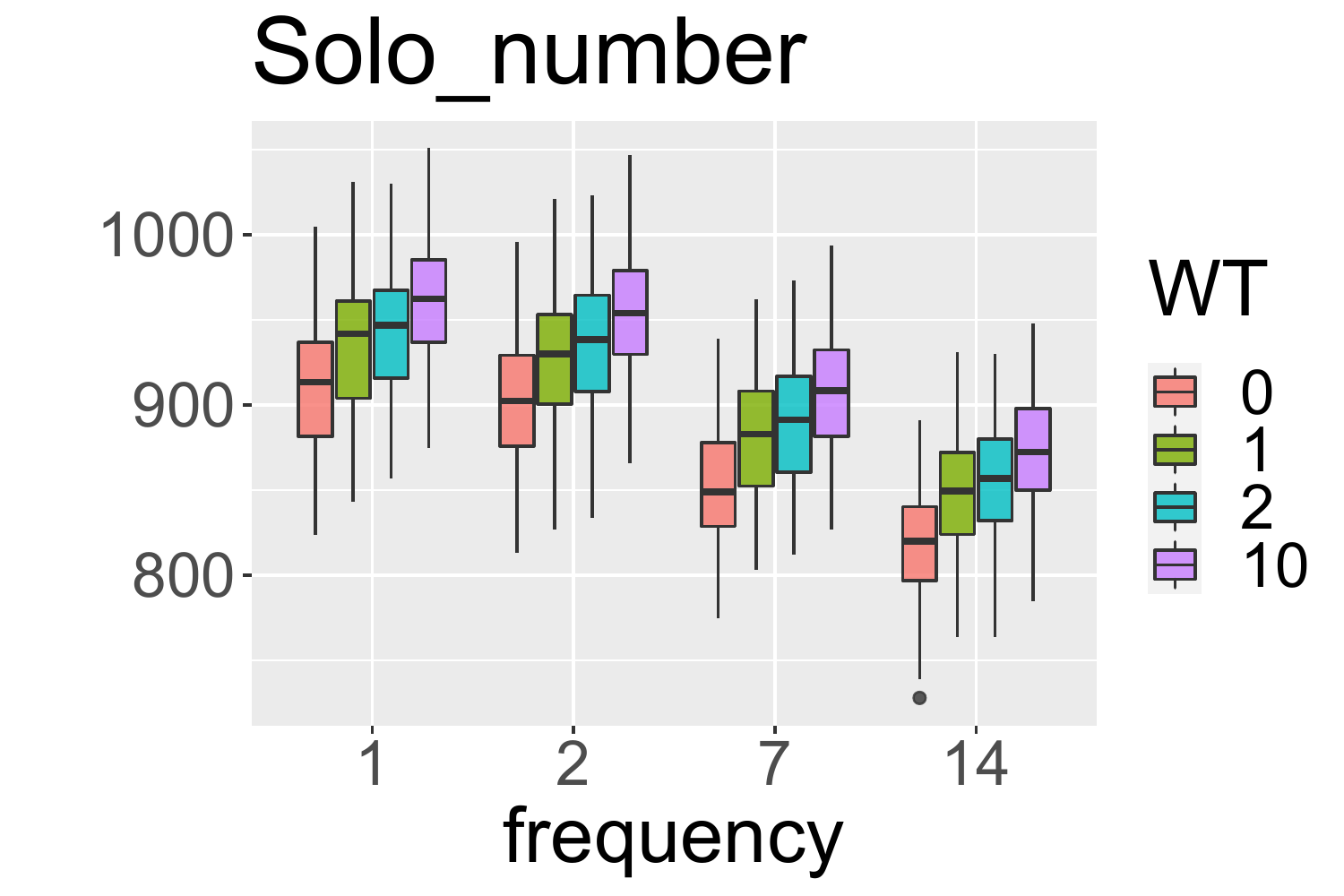}
	\end{minipage}
	\hspace{1cm}
	\begin{minipage}{.4\textwidth}
		\includegraphics[width=1.2\textwidth]{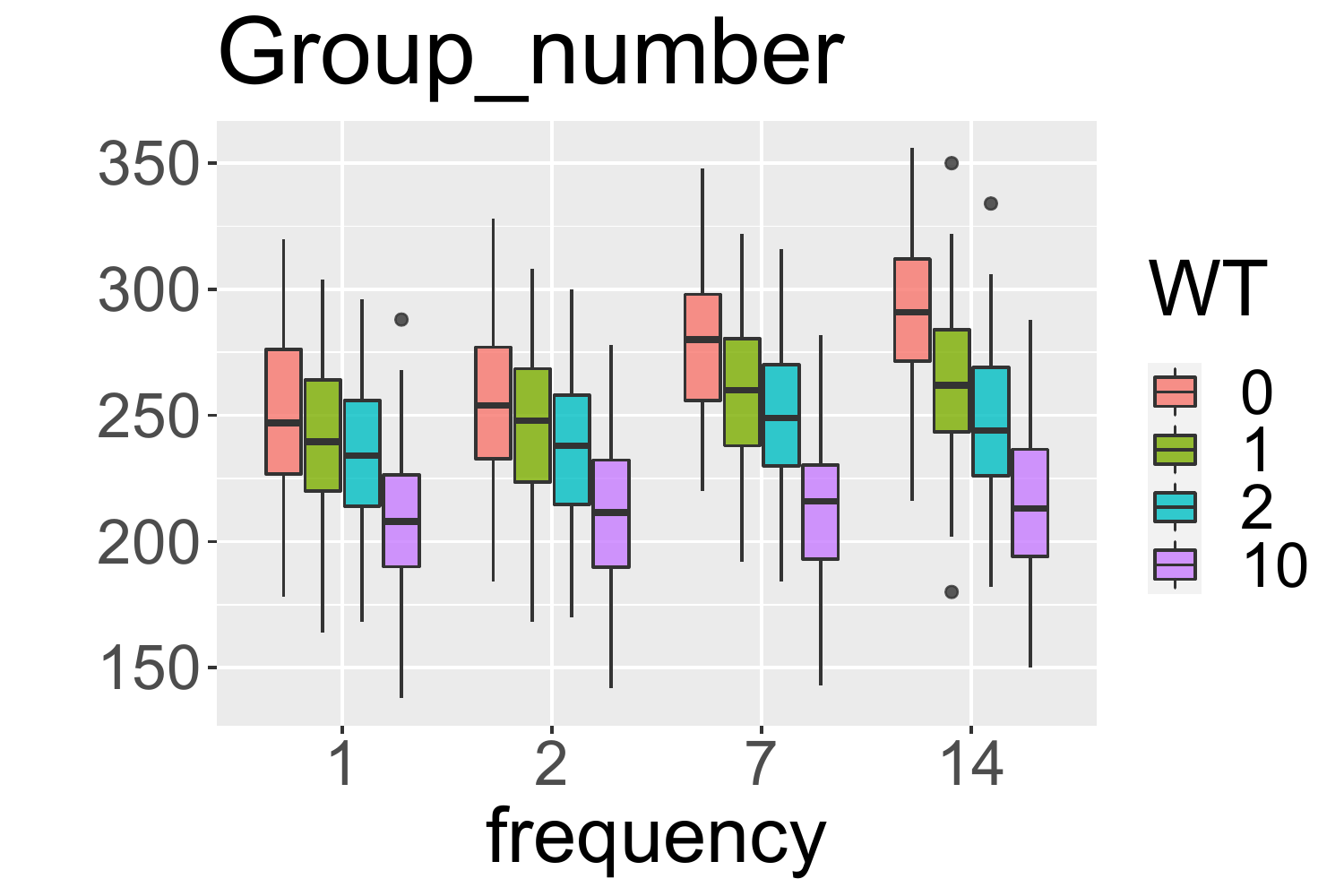}
	\end{minipage}
	\caption{The Solo and Group results with  0,1,2,10 weight on the waiting time} 
	\label{fig:boxplots_wait_all_group}
\end{figure}

Figure \ref{fig:boxplots_wait_all_group} presents the Solo-groups' and groups' distributions of the results.

With the increase in the weight of the waiting time priority, the number of pairs increased, but the number of groups decreased.  We can observe a similar trend for each match-frequency, hence giving priority for the waiting time is good for forming more pairs. 

\begin{figure}[h!]

	\begin{minipage}{.4\textwidth}
		\includegraphics[width=1.2\textwidth]{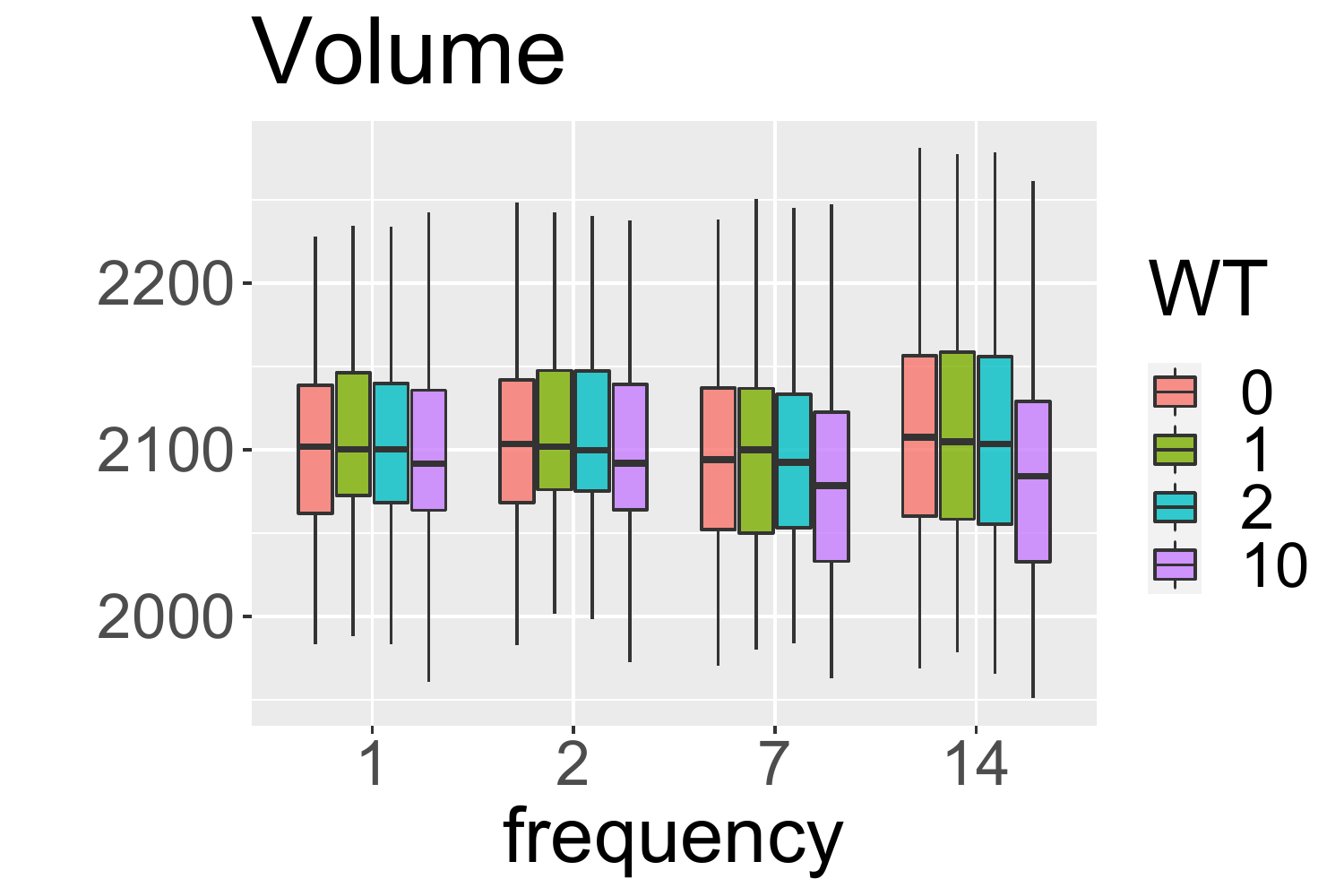}
	\end{minipage}
	\hspace{1cm}
	\begin{minipage}{.4\textwidth}
		\includegraphics[width=1.2\textwidth]{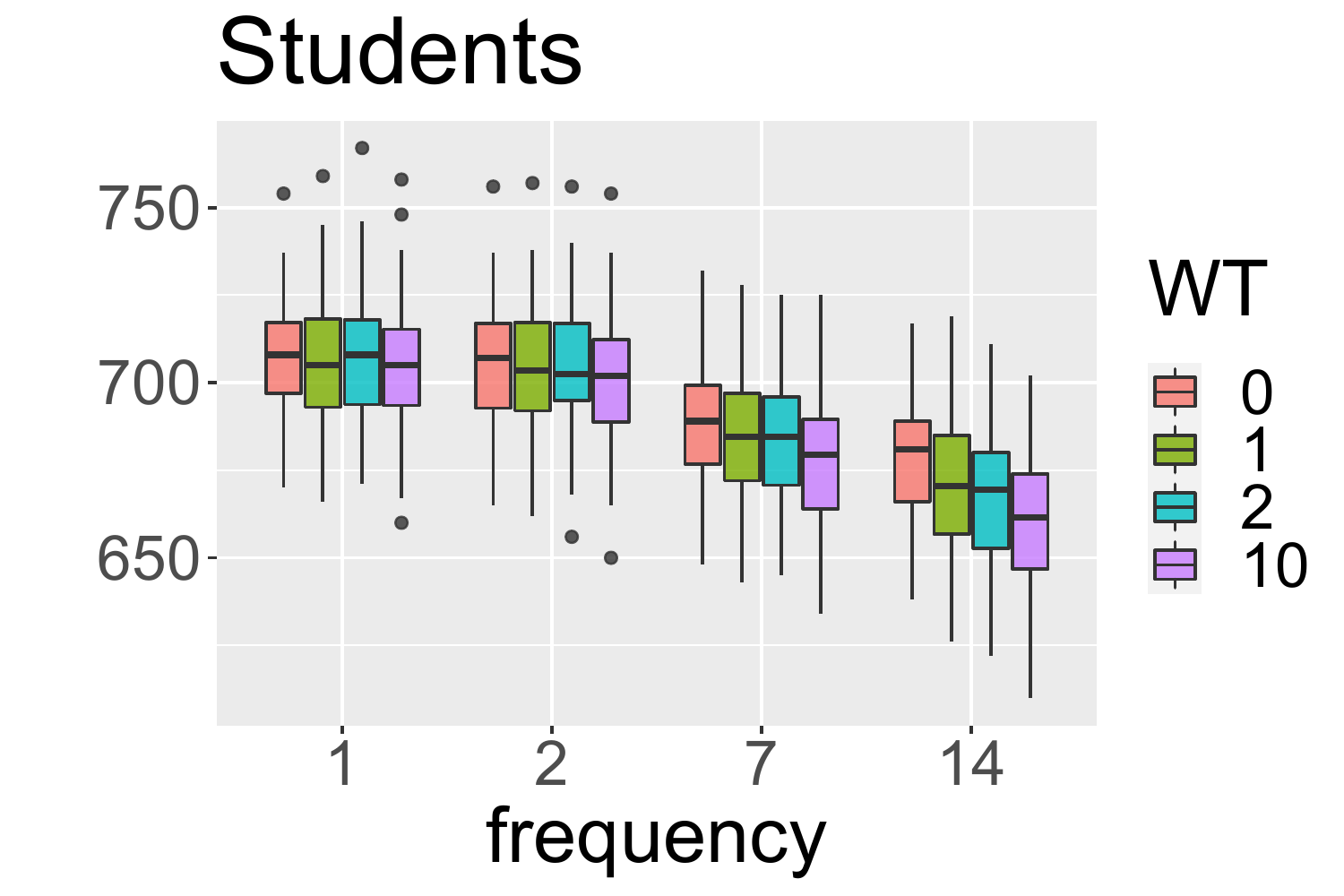}
	\end{minipage}
	\caption{The Volume and Number of students results with  0,1,2,10 weight on the waiting time} 
	\label{fig:boxplots_wait_quantity}
\end{figure}

However, overall, we can notice a small decrease in the Volume, as Figure \ref{fig:boxplots_wait_quantity} presents. For WT=1 and 2 this reduction is negligible, but for WT=10 it is significant. Thus giving priority by the waiting time, we loose more mentoring time with the decreased number of groups than what we gain with more pairs.

The same figure presents the change in the number of students as well. In every run-frequency, a higher WT resulted in a decrease in the number of students allocated. 

\begin{figure}[h!]

		\begin{minipage}{.4\textwidth}
		\includegraphics[width=1.2\textwidth]{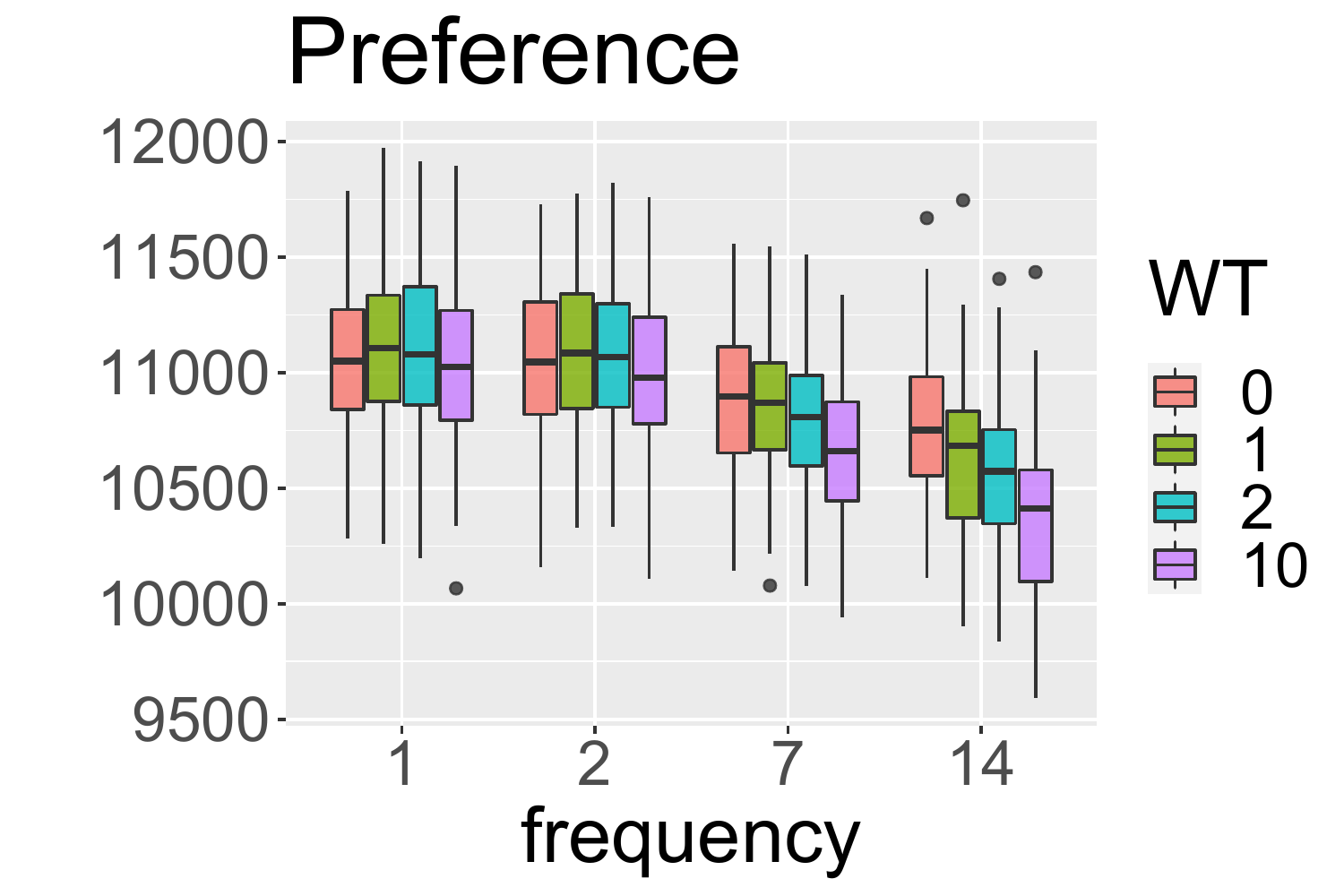}
	\end{minipage}
	\hspace{1cm}
	\begin{minipage}{.4\textwidth}
		\includegraphics[width=1.2\textwidth]{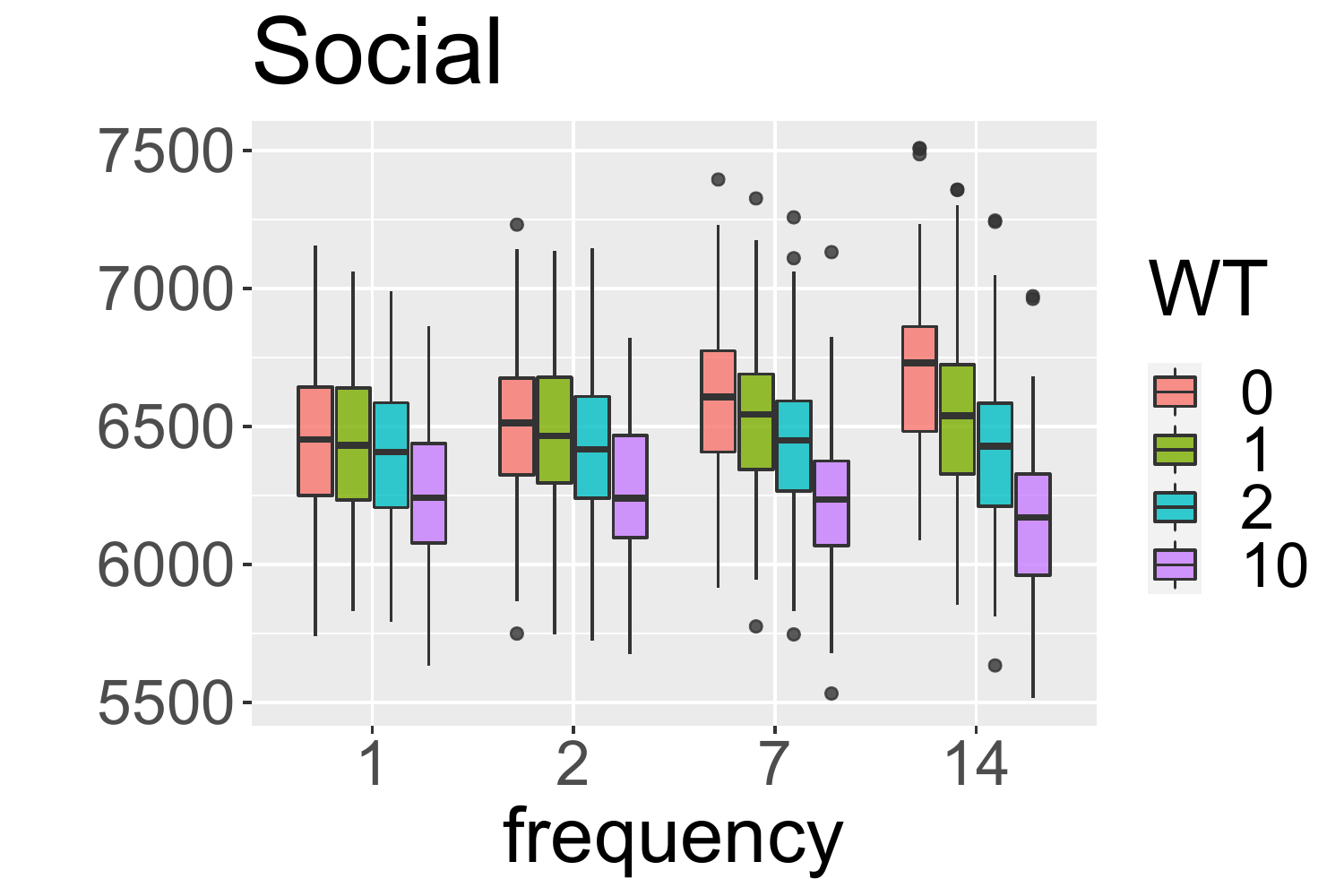}
	\end{minipage}
	\caption{The Preference and Social results with  0,1,2,10 weight on the waiting time} 
	\label{fig:boxplots_wait_quality}
\end{figure}

Figure \ref{fig:boxplots_wait_quality} presents the Preference and Social scores. In both cases, higher WT resulted in a relapse. It is connected to the decrease in Volume since fewer mentoring hours in general means worse Preference and Social scores.

Hence the prioritisation by the waiting time increase only the number of pairs and decrease many other aspects of the programme. The setback in these values is caused by the multiple demands and offers. For example, a student may have two different subjects requested. If there is a match for her first subject then she remains in the programme with her second remaining subject with an extended duration of stay, and so in a later run she may well receive a mentor for her second subject as well due to her increased priority instead of allocating a freshly registered student for her first subject. In fact, the freshly registered student can even have higher Social score, or other scores can also be better for her, but the high waiting time priority for the aforementioned student overrules these scores. Thus the freshly registered student may not even get a mentor, whilst the earlier matched student will get multiple mentors. Therefore there is a setback in both match quality and somewhat also in quantity.  

\begin{figure}[h!]
	\centering
	\captionsetup{justification=centering, margin = 2cm}
	\begin{minipage}{.4\textwidth}
		\includegraphics[width=1.2\textwidth]{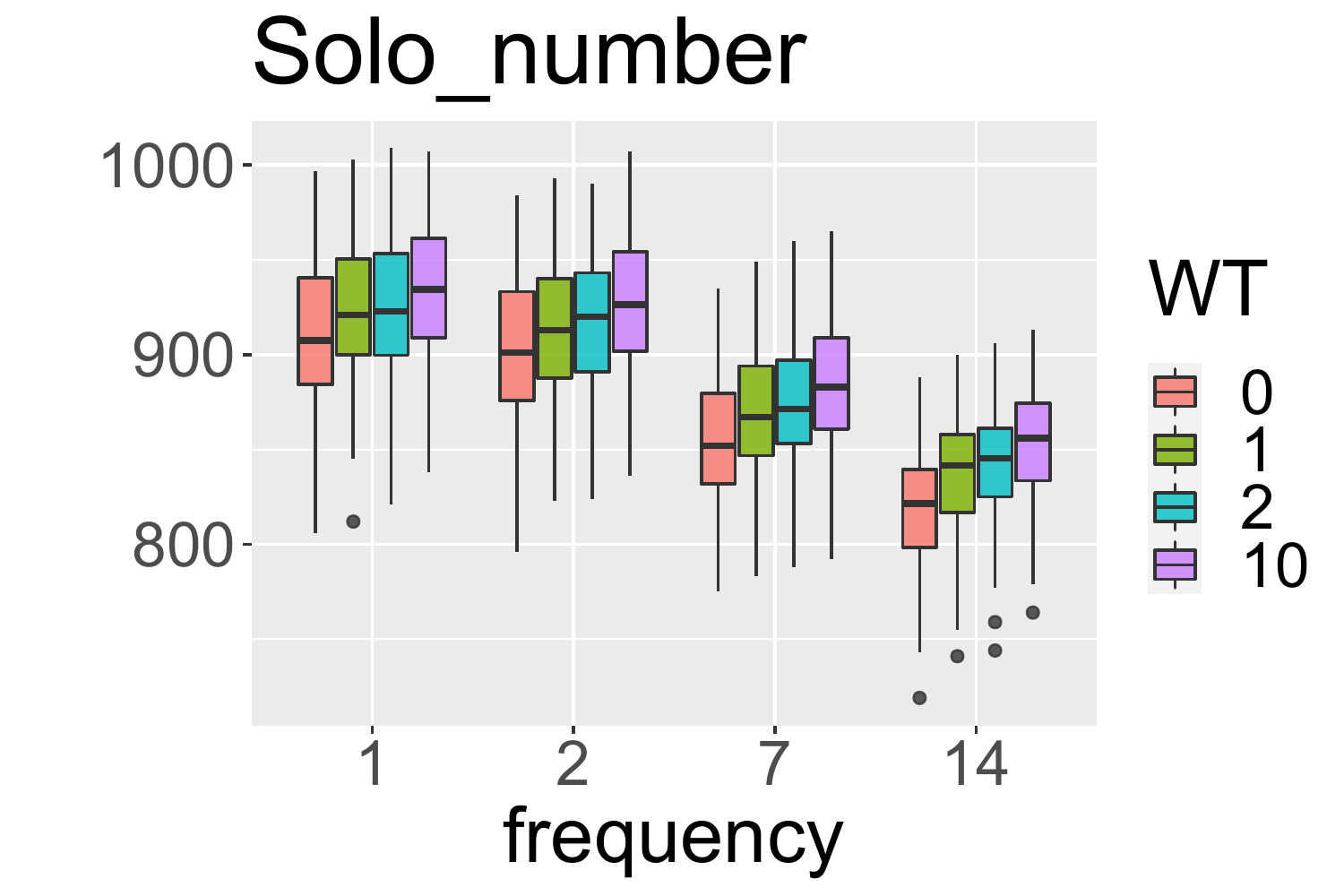}
	\end{minipage}
	\hspace{1cm}
	\begin{minipage}{.4\textwidth}
		\includegraphics[width=1.2\textwidth]{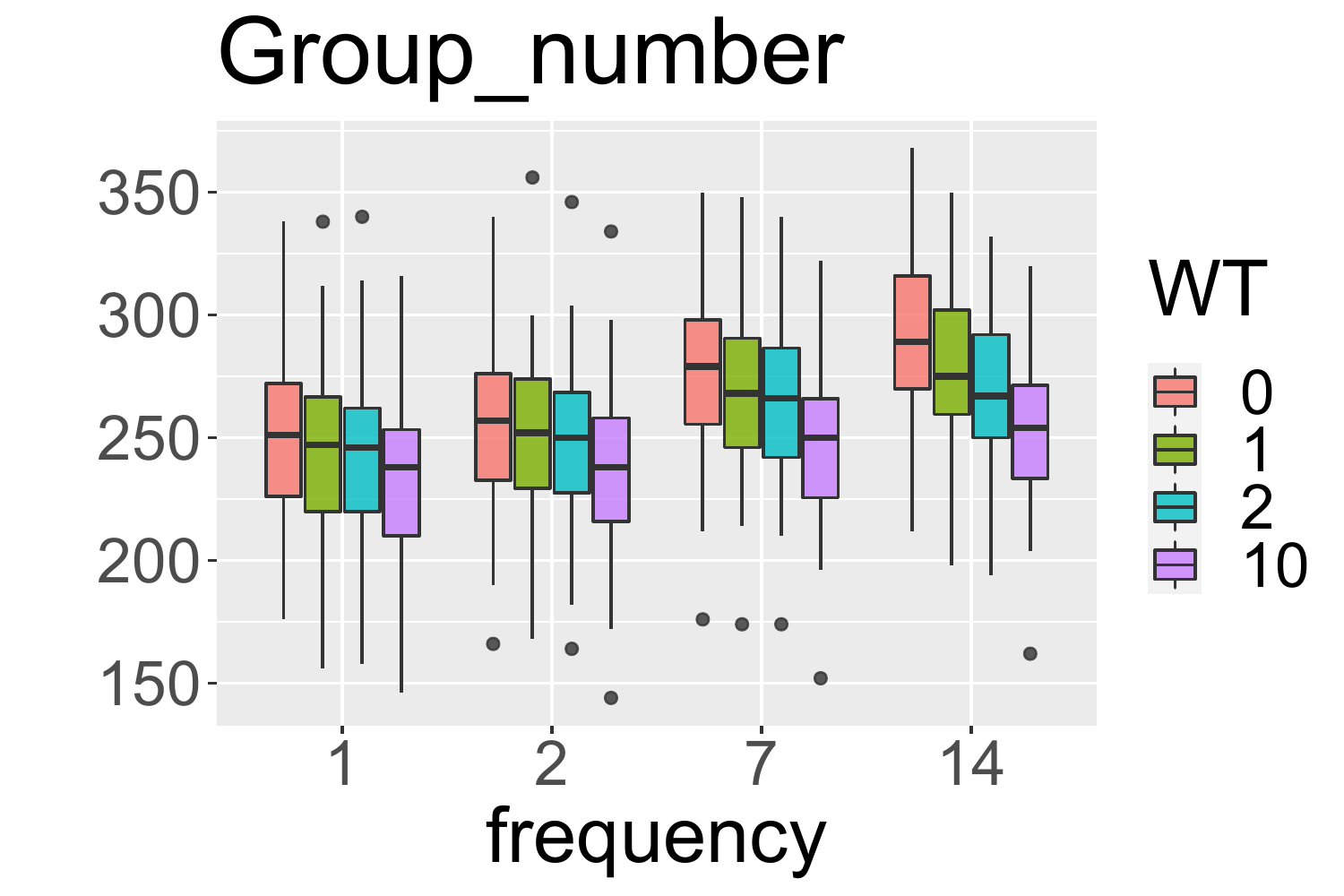}
	\end{minipage}
	%\vspace{0pt}
	\begin{minipage}{.4\textwidth}
		\includegraphics[width=1.2\textwidth]{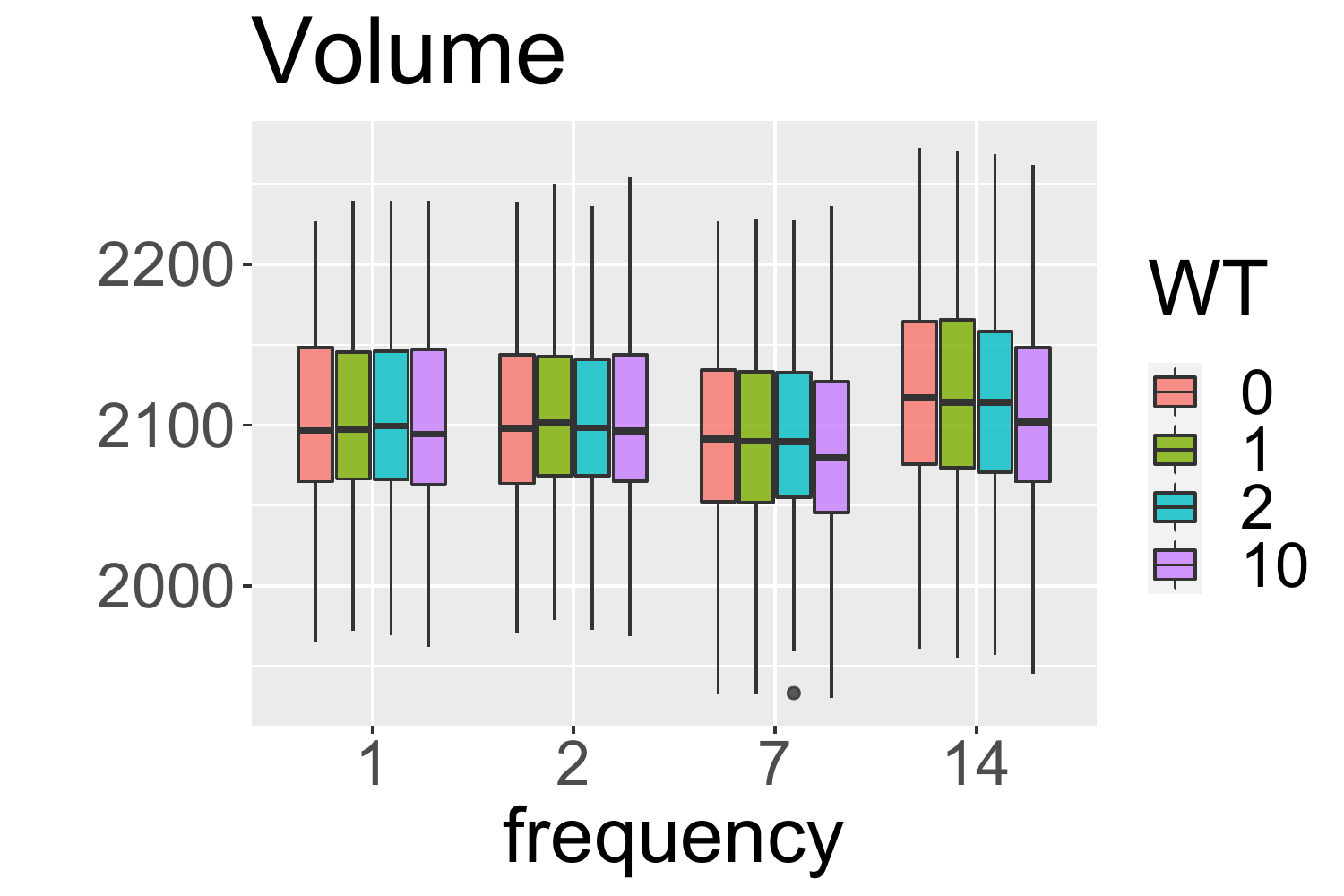}
	\end{minipage}
	\hspace{1cm}
	\begin{minipage}{.4\textwidth}
		\includegraphics[width=1.2\textwidth]{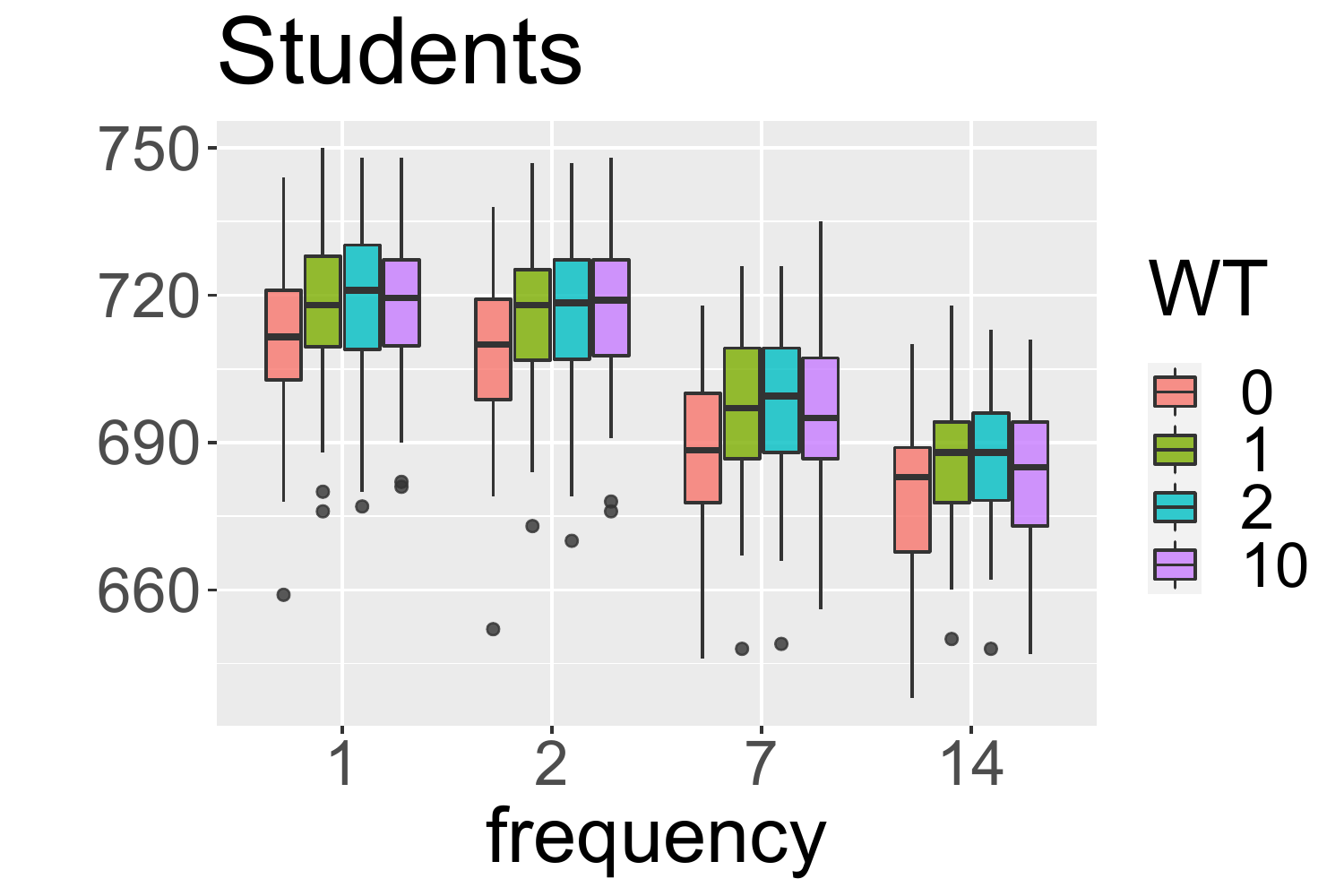}
	\end{minipage}
		\begin{minipage}{.4\textwidth}
		\includegraphics[width=1.2\textwidth]{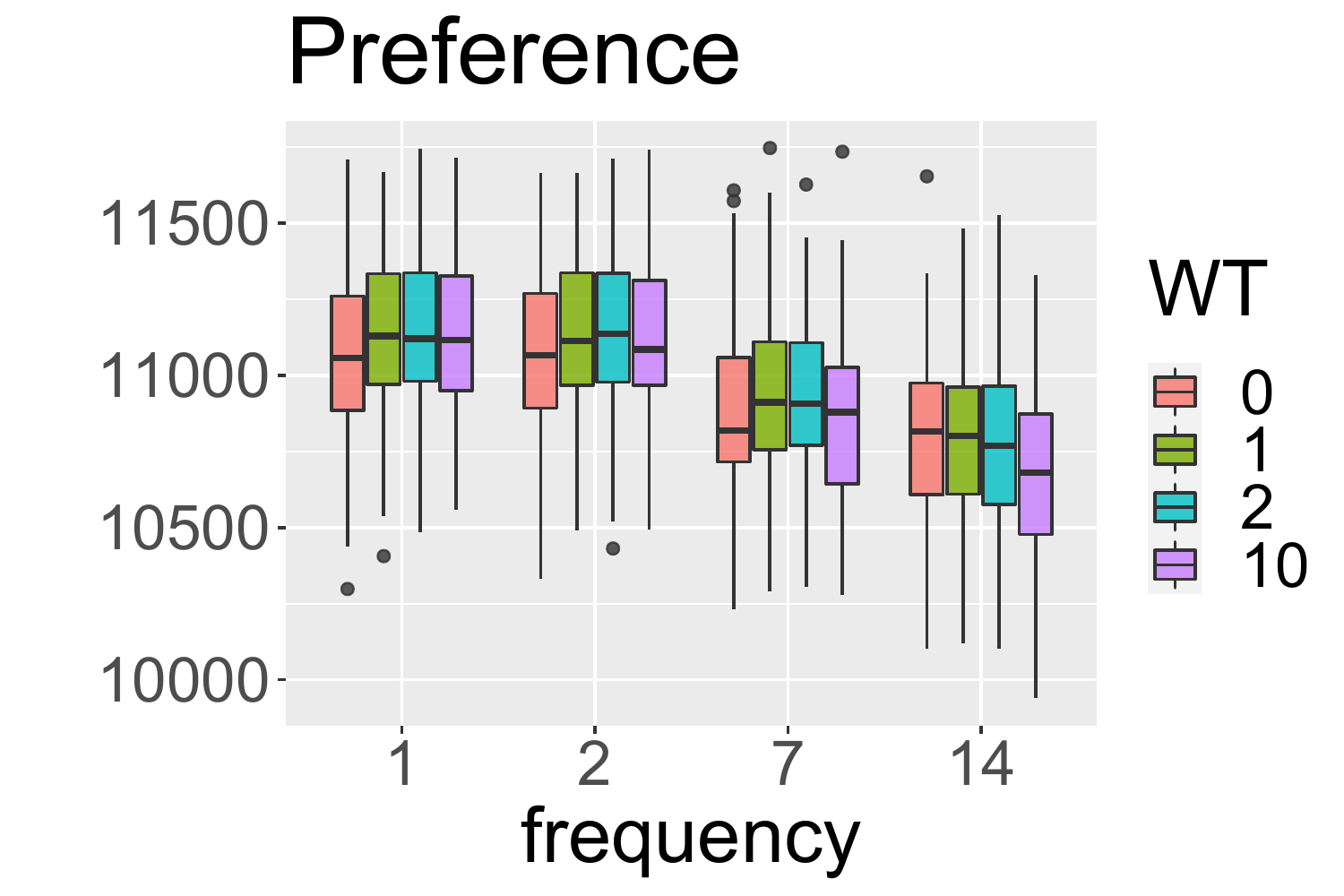}
	\end{minipage}
	\hspace{1cm}
	\begin{minipage}{.4\textwidth}
		\includegraphics[width=1.2\textwidth]{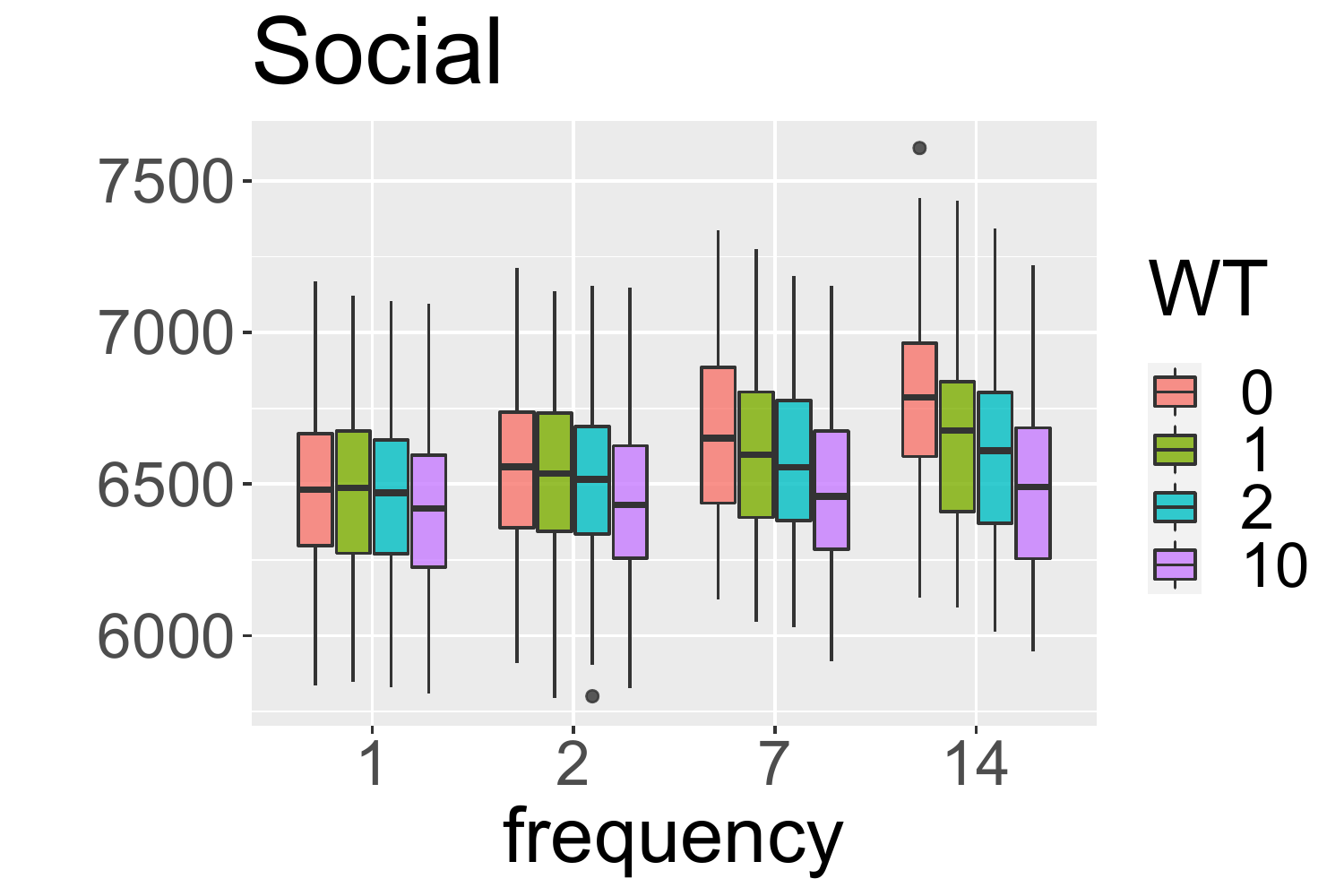}
	\end{minipage}
	\caption{The results with weights 0,1,2,10 for the waiting time regarding the first preferred subjects only} 
	\label{fig:boxplots_wait_1}
\end{figure}

To reduce the setback, caused by the multiple demands and offers by the members, we also investigated how the solution would change, if we only consider the extra priority on the waiting time for the first preferred subjects of the students. Figure \ref{fig:boxplots_wait_1} presents the results of these models, with the same setups as earlier.

In the pairs and groups, the effect was similar to the earlier investigated case. The weight on the waiting time of the first subject also increased the number of pairs and decreased the number of groups. However, the change is noticeably smaller. Therefore the Volume appears to be more stable, but still has a slight reduction as the WT increases. 
However, the trend in the number of students changed. In this case, a higher weight on the waiting time of the first subject increased the
number of students matched. Because each student has one subject with this weight, those students who have not been selected yet has a higher chance to get matched than before. %On the other side,  because only one subject has weight, it does not take away the opportunity from the other students that much. Hence there is a small increase.

Regarding the preferences, we can notice a small increase when the runs occur in every day or in every two days. However, the score decreases when the runs happen less frequently. The reason behind the change in the trend is maybe because of the reduction in the number of groups.  In general, more frequent runs resulted in more pairs and fewer number of groups. The weights for the waiting time also increased the number of pairs and reduced the possibility to form a group. More pairs with higher preferences increased the preference score, but having fewer groups decreased it. 

The Social score decreased for every run frequency as we increased the WT. However, the effect is much smaller compared to the case when every subject was weighted. %It is less when WT is smaller, and the runs occur more frequently.

\section{Conclusion}

In this paper we have described the optimisation aspect of a joint NGO project for allocating voluntary mentors to students. By taking the participants' preferences into account we aimed to create desirable pairs and study groups by using integer programming techniques for solving the dynamic allocation problem in the real application and also for generated data. We believe that the lessons learned can be useful for other countries and for similar applications. 

As a future work we are going to conduct a follow-up analyses on the data that we are now gathering in the Hungarian application since the latest lockdown for secondary schools in the 2020/2021 winter and spring period.

\section*{Acknowledgement}

We thank our colleagues, D\'aniel Horn and Zolt\'an Hermann for helping designing the application on behalf of our Institute, D\'avid Burka and his company for developing the website of the application (onkentesmentoralas.hu), and our collaborators in the two partner organisations, the Hungarian Reformat Church Aid and \#school. Bir\'o and Gyetvai are supported by the Hungarian Scientific Research Fund -- OTKA (no.\ K129086).

\bibliographystyle{plain}
\bibliography{OR-matching}

\begin{thebibliography}{10}

\bibitem{Abraham2007}
David~J. Abraham, Avrim Blum, and Tuomas Sandholm.
\newblock Clearing algorithms for barter exchange markets: Enabling nationwide
  kidney exchanges.
\newblock In {\em Proceedings of the 8th ACM Conference on Electronic
  Commerce}, EC '07, pages 295--304, New York, NY, USA, 2007. ACM.

\bibitem{agarwal2019market}
Nikhil Agarwal, Itai Ashlagi, Eduardo Azevedo, Clayton~R Featherstone, and
  {\"O}mer Karaduman.
\newblock Market failure in kidney exchange.
\newblock {\em American Economic Review}, 109(11):4026--70, 2019.

\bibitem{Agarwal2020}
Nikhil Agarwal, Itai Ashlagi, M~Rees, P~Somaini, and D~Waldinger.
\newblock Equilibrium allocations under alternative waitlist designs: Evidence
  from deceased donor kidneys.
\newblock {\em Econometrica, forthcoming}, 2020.

\bibitem{ABMcB2016}
Kolos~Csaba {\'A}goston, P{\'e}ter Bir{\'o}, and Iain McBride.
\newblock Integer programming methods for special college admissions problems.
\newblock {\em Journal of Combinatorial Optimization}, 32(4):1371--1399, 2016.

\bibitem{ABSz2018}
Kolos~Csaba {\'A}goston, P{\'e}ter Bir{\'o}, and Rich{\'a}rd Sz{\'a}nt{\'o}.
\newblock Stable project allocation under distributional constraints.
\newblock {\em Operations Research Perspectives}, 5:59--68, 2018.

\bibitem{andersson2018dynamic}
Tommy Andersson, Lars Ehlers, and Alessandro Martinello.
\newblock Dynamic refugee matching.
\newblock 2018.

\bibitem{AshlagiRoth2020}
Itai Ashlagi and Alvin~E Roth.
\newblock Kidney exchange: an operations perspective.
\newblock Working paper, 2020.

\bibitem{bansak2018improving}
Kirk Bansak, Jeremy Ferwerda, Jens Hainmueller, Andrea Dillon, Dominik
  Hangartner, Duncan Lawrence, and Jeremy Weinstein.
\newblock Improving refugee integration through data-driven algorithmic
  assignment.
\newblock {\em Science}, 359(6373):325--329, 2018.

\bibitem{biro2017applications}
P{\'e}ter Bir{\'o}.
\newblock Applications of matching models under preferences.
\newblock 2017.

\bibitem{biroetal2019a}
P{\'e}ter Bir{\'o}, Bernadette Haase-Kromwijk, Joris van~de Klundert, and
  et~al.
\newblock Building kidney exchange programmes in {E}urope -- an overview of
  exchange practice and activities.
\newblock {\em Transplantation}, 103(7):1514--1522, 2019a.

\bibitem{BMMcB2014}
P{\'e}ter Bir{\'o}, David~F Manlove, and Iain McBride.
\newblock The hospitals/residents problem with couples: Complexity and integer
  programming models.
\newblock In {\em International Symposium on Experimental Algorithms}, pages
  10--21. Springer, 2014.

\bibitem{Biro2014}
P{\'e}ter Bir{\'o} and Eric McDermid.
\newblock Matching with sizes (or scheduling with processing set restrictions).
\newblock {\em Discrete Applied Mathematics}, 164:61--67, 2014.

\bibitem{biroetal2019b}
P\'eter Bir\'o, Joris {van de Klundert}, David Manlove, William Pettersson,
  Tommy Andersson, Lisa Burnapp, Pavel Chromy, Pablo Delgado, Piotr Dworczak,
  Bernadette Haase, Aline Hemke, Rachel Johnson, Xenia Klimentova, Dirk
  Kuypers, Alessandro {Nanni Costa}, Bart Smeulders, Frits Spieksma, María~O.
  Valentín, and Ana Viana.
\newblock Modelling and optimisation in european kidney exchange programmes.
\newblock {\em European Journal of Operational Research}, 2019b.

\bibitem{bloch2020matching}
Francis Bloch, David Cantala, and Dami{\'a}n Gibaja.
\newblock Matching through institutions.
\newblock {\em Games and Economic Behavior}, 2020.

\bibitem{budish2017course}
Eric Budish, G{\'e}rard~P Cachon, Judd~B Kessler, and Abraham Othman.
\newblock Course match: A large-scale implementation of approximate competitive
  equilibrium from equal incomes for combinatorial allocation.
\newblock {\em Operations Research}, 65(2):314--336, 2017.

\bibitem{Cao2010}
Nguyen~Vi Cao, Emmanuel Fragniere, Jacques-Antoine Gauthier, Marlene Sapin, and
  Eric~D Widmer.
\newblock Optimizing the marriage market: An application of the linear
  assignment model.
\newblock {\em European Journal of Operational Research}, 202(2):547--553,
  2010.

\bibitem{Delormeetal2019}
Maxence Delorme, Sergio Garc{\'\i}a, Jacek Gondzio, J{\"o}rg Kalcsics, David
  Manlove, and William Pettersson.
\newblock Mathematical models for stable matching problems with ties and
  incomplete lists.
\newblock {\em European Journal of Operational Research}, 277(2):426--441,
  2019.

\bibitem{gale1962college}
David Gale and Lloyd~S Shapley.
\newblock College admissions and the stability of marriage.
\newblock {\em The American Mathematical Monthly}, 69(1):9--15, 1962.

\bibitem{garg2010assigning}
Naveen Garg, Telikepalli Kavitha, Amit Kumar, Kurt Mehlhorn, and Juli{\'a}n
  Mestre.
\newblock Assigning papers to referees.
\newblock {\em Algorithmica}, 58(1):119--136, 2010.

\bibitem{Gerding2019}
Enrico Gerding, Alvaro Perez-Diaz, Haris Aziz, Serge Gaspers, Antonia Marcu,
  Nicholas Mattei, and Toby Walsh.
\newblock Fair online allocation of perishable goods and its application to
  electric vehicle charging.
\newblock 2019.

\bibitem{Harris2020}
Douglas~N Harris, Lihan Liu, Daniel Oliver, Cathy Balfe, Sara Slaughter, and
  Nicholas Mattei.
\newblock How america’s schools responded to the covid crisis.
\newblock {\em National Center for Research on Education Access and Choice \&
  Education Research Alliance for New Orleans.
  https://educationresearchalliancenola.
  org/files/publications/20200713-Technical-Report-Harris-etal-How-Americas-Schools-Responded-to-the-COVID-Crisis.
  pdf}, 2020.

\bibitem{KM2014}
Augustine Kwanashie and David~F Manlove.
\newblock An integer programming approach to the hospitals/residents problem
  with ties.
\newblock In {\em Operations Research Proceedings 2013}, pages 263--269.
  Springer, 2014.

\bibitem{leshno2019dynamic}
Jacob Leshno.
\newblock Dynamic matching in overloaded waiting lists.
\newblock {\em Available at SSRN 2967011}, 2019.

\bibitem{Manlove2013book}
David Manlove.
\newblock {\em Algorithmics of matching under preferences}, volume~2.
\newblock World Scientific, 2013.

\bibitem{Mattei2018}
Nicholas Mattei, Abdallah Saffidine, and Toby Walsh.
\newblock Fairness in deceased organ matching.
\newblock In {\em Proceedings of the 2018 AAAI/ACM Conference on AI, Ethics,
  and Society}, pages 236--242, 2018.

\bibitem{prendergast2016allocation}
Canice Prendergast.
\newblock The allocation of food to food banks.
\newblock {\em EAI Endorsed Trans. Serious Games}, 3(10):e4, 2016.

\bibitem{Roth2007}
Alvin~E. Roth, Tayfun S{\"o}nmez, and M.~Utku {\"U}nver.
\newblock Efficient kidney exchange: Coincidence of wants in markets with
  compatibility-based preferences.
\newblock {\em American Economic Review}, 97(3):828--851, June 2007.

\bibitem{So2016}
Mee~Chi So, Lyn~C Thomas, and Bo~Huang.
\newblock Lending decisions with limits on capital available: The polygamous
  marriage problem.
\newblock {\em European Journal of Operational Research}, 249(2):407--416,
  2016.

\bibitem{trapp2018placement}
Andrew~C Trapp, Alexander Teytelboym, Alessandro Martinello, Tommy Andersson,
  Narges Ahani, et~al.
\newblock Placement optimization in refugee resettlement.
\newblock Technical report, 2018.

\end{thebibliography}

\end{document}